\documentclass[aps,prd, nofootinbib,superscriptaddress, article]{revtex4}

\usepackage{amsmath, amsfonts, amsthm, amssymb, graphicx, slashed}
\usepackage{subfig}
\usepackage{epstopdf}

\def\be{\begin{equation}}
\def\ee{\end{equation}}
\def\ba{\begin{eqnarray}}
\def\ea{\end{eqnarray}}

\def\bdm{\begin{displaymath}}
\def\edm{\end{displaymath}}

\def\bq{\begin{quote}}
\def\eq{\end{quote}}

\def\del{\partial}
\def\ltap{\ \raise.3ex\hbox{$<$\kern-.75em\lower1ex\hbox{$\sim$}}\ }
\def\gtap{\ \raise.3ex\hbox{$>$\kern-.75em\lower1ex\hbox{$\sim$}}\ }
\def\gl{\ \raise.5ex\hbox{$>$}\kern-.8em\lower.5ex\hbox{$<$}\ }
\def\roughly#1{\raise.3ex\hbox{$#1$\kern-.75em\lower1ex\hbox{$\sim$}}}

 at 10truept

\def\GB{{\hat{\cal{G}}}}

\def\half{\textstyle{1\over2}}

\def\ons{{\it on-shell-in-a }}

\newcommand{\beq}{\begin{equation}}
\newcommand{\eeq}{\end{equation}}
\newcommand{\bea}{\begin{eqnarray}}
\newcommand{\eea}{\end{eqnarray}}
\newcommand{\beqa}{\begin{eqnarray}}
\newcommand{\eeqa}{\end{eqnarray}}

\newcommand{\dslash}{{\slashed \partial}}
\newcommand{\pslash}{{\slashed p}}

\begin{document}

\title{Self-tuning and the derivation of  the  Fab Four}

\author{Christos Charmousis} 
\affiliation{LPT, CNRS UMR 8627, Universit\'e Paris Sud-11, 91405 Orsay Cedex, France.}
\affiliation{LMPT, CNRS UMR 6083, Universit\'e Fran\c{c}ois Rabelais-Tours, 37200, France} 
\author{Edmund J. Copeland} 
\author{Antonio Padilla} 
\author{Paul M. Saffin} 
\affiliation{School of Physics and Astronomy, 
University of Nottingham, Nottingham NG7 2RD, UK} 

\date{\today}

\begin{abstract}
We have recently proposed a special class of scalar tensor theories known as {\it the Fab Four}.  These arose from attempts to  analyse the  cosmological constant problem  within the context of Horndeski's most general scalar tensor theory.  {\it The Fab Four} together give rise to a model of self-tuning,  with the relevant solutions evading Weinberg's no-go theorem by relaxing the condition of Poincar\'e invariance in the scalar sector.  {\it The Fab Four}  are made up of four geometric terms in the action with each term containing a free potential function of the scalar field.  In this paper we rigorously derive this model from the general model of Horndeski, proving that {\it the Fab Four} represents  the {\it only} classical scalar tensor theory of this type that has any hope of tackling the cosmological constant problem. We present the full equations of motion for this theory, and give an heuristic argument to suggest that one might be able to keep radiative corrections under control. We also give  {\it the  Fab Four} in terms of the potentials presented in  Deffayet {\it et al}'s version of Horndeski. 
\end{abstract}


\maketitle

\section{Introduction}

The cosmological constant problem has been 
described as the most embarrassing fine-tuning problem in Physics today. According to our current understanding of particle physics and effective {quantum} field theory, the vacuum receives zero point energy contributions from each particle species right up to the {UV} cut-off, which may be as high as the Planck scale. 
The trouble is that in General Relativity, any matter, including vacuum energy, gravitates and  the only way to make it compatible with observation is to demand  considerable fine-tuning between the vacuum energy and the bare cosmological constant.  The situation is exacerbated by phase transitions in the early universe that can give rise to constant shifts in the vacuum energy contribution.  To date, particle physicists have failed to come up with a satisfactory solution to this problem, so some recent attempts have instead focussed on gravitational physics. This  alternative approach requires a non-trivial  modification  of Einstein's theory at large distances (see \cite{review} for a detailed review of modified gravity).

One particularly interesting direction involves scalar-tensor theories of gravity. 
It seems sensible to require that any theory maintains second order field equations in order to avoid an Ostrogradski instability \cite{ostro}, and the most general scalar-tensor theory satisfying that criteria  in four dimensions was written down back in 1974 by Horndeski  \cite{horndeski:1974} (it has recently been rediscovered independently in \cite{general}). Such theories of modified gravity cover a wide range of models, 
ranging from Brans-Dicke gravity \cite{bdgravity}  to the recent models \cite{covgal,galmodels} inspired by galileon theory \cite{galileon}. 
Galileon models are examples of  higher order scalar tensor Lagrangians with second order field equations, and, as a result, they are closely related to Kaluza-Klein compactifications of higher dimensional Lovelock theories \cite{kkl, VanAcoleyen:2011mj}.  Of course all of these scalar-tensor models  can be considered as special cases of Horndeski's original action. 


%
%

 In \cite{Charmousis:2011bf} we obtained a new class of solutions arising out of Horndeski's theory on FLRW backgrounds. The new solutions gave a viable self-tuning mechanism for solving the (old) cosmological constant problem, at least at the classical level, by completely screening the spacetime curvature from the net cosmological constant. This would seem to be in violation of Weinberg's famous no-go theorem \cite{nogo} that forbids precisely this kind of self-adjustment mechanism. However, Weinberg assumes Poincar\'e invariance to hold universally across all fields whereas we allow it to be broken in the scalar field sector.  In other words, we continue to require  Poincar\'e invariance at the level of   spacetime curvature, but not  at the level of the self-adjusting scalar field.   A similar approach was adopted in the context of bigalileon theory  \cite{bigalileon} where only a small vacuum energy could be successfully screened away. In \cite{Charmousis:2011bf}, we provided a brief sketch of how the system works for  scalar tensor theories where matter is only minimally coupled to the metric (required to ensure compatibility with Einstein's Equivalence Principle (EEP)). By demanding the presence of a viable  self-tuning mechanism we were able to place powerful restrictions on the allowed form of Horndeski's original Lagrangian. Whereas the  original model is complicated, with many arbitrary functions of both the scalar and its derivatives, we showed that  once the model is passed through our self-tuning filter (to be defined shortly), it reduces in form to just four base Lagrangians each depending on an arbitrary function of the scalar only, coupled to a curvature term. We called these base Lagrangians {\it the Fab Four}: John, Paul, George and Ringo. 
 
 Together, {\it the Fab Four} make up the most general scalar-tensor theory capable of self-tuning. Individually they are given by the following
  \begin{eqnarray}
\label{eq:john}
{\cal L}_{john} &=& \sqrt{-g} V_{john}(\phi)G^{\mu\nu} \nabla_\mu\phi \nabla_\nu \phi \\
\label{eq:paul}
{\cal L}_{paul} &=&\sqrt{-g}V_{paul}(\phi)   P^{\mu\nu\alpha \beta} \nabla_\mu \phi \nabla_\alpha \phi \nabla_\nu \nabla_\beta \phi \\
\label{eq:george}
{\cal L}_{george} &=&\sqrt{-g}V_{george}(\phi) R \\
\label{eq:ringo}
{\cal L}_{ringo} &=& \sqrt{-g}V_{ringo}(\phi) \GB
\end{eqnarray}
where $R$ is the Ricci scalar, $G_{\mu\nu}$ is the Einstein tensor, $P_{\mu\nu\alpha \beta}$ is the double dual of the Riemann tensor \cite{mtw}, $\GB=R^{\mu\nu \alpha \beta} R_{\mu\nu \alpha \beta}-4R^{\mu\nu}R_{\mu\nu}+R^2$ is the Gauss-Bonnet combination, and in what follows the Greek indices $\mu,\nu =0..3$. The purpose of this paper is  to rigorously derive the conditions that lead to these four base Lagrangians, showing how they naturally lead to self-tuning solutions, provided that $\{V_{john}, V_{paul}, V_{george}\} \neq \{0,0, constant\}$. Note that this constraint means that GR is {\it not} a {\it Fab Four} theory, consistent with the fact that it does not have self-tuning solutions. 

To be clear as to what is meant by ``self-tuning", let us define our self-tuning filter. We require that
\begin{itemize}
\item the theory should admit a Minkowski vacuum\footnote{For simplicity throughout the introductory part of the text we have simply written, ``Minkowski vacuum" to stand for``a patch of Minkowski vacuum". This technical issue will be made clear later on in section \ref{self-tune}} for any value of the net cosmological constant
\item this should remain true before and after any phase transition where the cosmological constant jumps instantaneously by a finite amount.
\item the theory should permit a non-trivial cosmology
\end{itemize}
The last condition ensures that Minkowski space is not the only cosmological solution available, something that is certainly required by observation. The idea is that the cosmological field equations should be dynamical, with the Minkowski solution corresponding to some sort of fixed point. In other words, once we are on a Minkowski solution, we stay there -- otherwise we evolve to it dynamically. This last statement would indicate that the self-tuning vacuum is an attractive fixed point. We do not prove this here, but in our companion paper on cosmology \cite{fab4cos} we will see plenty of examples where it is indeed the case.

The first two conditions are the basic requirements of any successful self-tuning mechanism. There are many examples in the literature which pass the first condition, but fall down at the second. This includes the much explored co-dimension two braneworld models in which the compact extra dimensions are shaped like a rugby ball \cite{Carroll}. The  brane tension controls  the deficit angle, while the brane geometry  is completely determined by the bulk 
cosmological constant and the magnetic flux. Therefore, this passes our first condition.  However, when the brane tension changes after a phase transition it affects the 
brane curvature via the backdoor, by altering magnetic flux and the  theory falls foul of our second condition \cite{cline}. 

It is interesting to note that any diffeomorphism invariant theory that passes both the first and second condition will admit a Minkowski solution in the presence of {\it any} cosmological fluid, not just a cosmological constant.  The point is that our vacuum energy density corresponds to a piecewise constant function, with discontinuities at the phase transitions. In principle these transitions can occur at any given time, so a Minkowski solution can be returned for all piecewise constant energy densities. The energy density of  an arbitrary cosmological fluid can be well approximated by a piecewise constant function, and so it follows that  it must also admit a  Minkowski solution. Like we said, this property must hold for {\it any} diffeomorphism invariant  theory passing our first two conditions, and not just {\it the Fab Four}.  We might worry that this prevents any hope of a sensible matter dominated cosmology. However, this is where the third condition comes into play, and we once again refer the reader to our companion paper \cite{fab4cos} for evidence that sensible cosmologies are indeed possible within this theory.

Even so, the main aim of this paper is not to extoll the virtues of  {\it the Fab Four} but to push a very general class of modified gravity theories through our self-tuning filter and to see what happens. In a sense we are testing the scope of Weinberg's theorem, relaxing one of his assumptions and seeing how far we can go.  It turns  out that our filter is very efficient -- it removes most of Horndeski's original theory-- but it is not $100\%$ efficient. We are left with  {\it the Fab Four}.

 The layout of the paper is as follows: in section \ref{horndeski-action} we present the original action of Horndeski  \cite{horndeski:1974}, minimally coupled to matter, and derive the Hamiltonian and scalar field equations of motion for the system. In section \ref{self-tune} we demonstrate how a self-tuning solution can in principle be obtained by relaxing Weinberg's no-go theorem to allow the scalar field to evolve in time. This is followed in section \ref{self-tune-horndeski} with a derivation of the self-tuned Horndeski action, where we show how the initial complicated Lagrangian reduces to four simple terms each one being an arbitrary function of the scalar field alone coupled to a curvature term. Of particular note is that any dependence on the kinetic energy of the scalar field drops out.  In section \ref{conc} we bring everything together and discuss further demands we may wish to make on our theory, over and above our original filter, ranging from cosmological and solar system tests, to issues of stability. We also elucidate the elegant geometrical structure possessed by {\it the Fab Four} and present their equations of motion in full. 
 
We have a number of appendices, most of which are technical additions to the main text. The exceptions are appendices \ref{app:dgsz} and \ref{radiative}. In  appendix \ref{app:dgsz} we present {\it the Fab Four} in the language of the  potentials of  Deffayet {\it et al}'s version of Horndeski \cite{general}.  In appendix \ref{radiative} we discuss the issue of radiative corrections to {\it the Fab Four}. This is an important question, because radiative corrections are at the heart of the cosmological constant problem.    We do not attempt a detailed analysis -- that is certainly beyond the scope of the current paper --- but we do perform some heuristic calculations. It seems that radiative corrections can be kept under control given some not too restrictive conditions.

\section{Horndeski's scalar-tensor theory}\label{horndeski-action}
The action we begin with for our general second-order scalar tensor theory is given by 
\be 
\label{eq:action}
S=S_{H}[g_{\mu\nu}, \phi]+S_m[g_{\mu\nu}; \Psi_n]
\ee
 where the Horndeski action, $S_{H}= \int d^4 x \sqrt{-g} {\cal L}_H$, is obtained from equation  (4.21) of \cite{horndeski:1974}, such that
 \ba
\label{eq:horndeskiFullLag}
{\cal L}_H&=& \kappa_1(\phi ,\rho)\delta^{\alpha \beta \gamma} _{\mu\nu\sigma}\nabla^\mu\nabla_\alpha \phi  R_{\beta \gamma} ^{\;\;\;\;\nu\sigma}
           -\frac{4}{3}\kappa_{1,\rho}(\phi ,\rho)\delta^{\alpha \beta \gamma} _{\mu\nu\sigma}\nabla^\mu\nabla_\alpha \phi\nabla^\nu\nabla_\beta \phi\nabla^\sigma\nabla_\gamma \phi \\\nonumber
        &~&+\kappa_3(\phi ,\rho)\delta^{\alpha \beta \gamma} _{\mu\nu\sigma}\nabla_\alpha \phi \nabla^\mu\phi  R_{\beta \gamma} ^{\;\;\;\;\nu\sigma}
           -4\kappa_{3,\rho}(\phi ,\rho)\delta^{\alpha \beta \gamma} _{\mu\nu\sigma}\nabla_\alpha \phi \nabla^\mu\phi \nabla^\nu\nabla_\beta \phi \nabla^\sigma\nabla_\gamma \phi \\\nonumber
        &~&+[F(\phi ,\rho)+2W(\phi )]\delta_{\mu\nu}^{\alpha \beta }R_{\alpha \beta }^{\;\;\;\;\mu\nu}
           -4F(\phi,\rho)_{,\rho}\delta_{\mu\nu}^{\alpha \beta }\nabla_\alpha \phi\nabla^\mu\phi \nabla^\nu\nabla_\beta \phi \\\nonumber
        &~&-3[2F(\phi ,\rho)_{,\phi }+4W(\phi )_{,\phi }+\rho\kappa_8(\phi,\rho)]\nabla_\mu\nabla^\mu\phi 
           +2\kappa_8\delta_{\mu\nu}^{\alpha \beta }\nabla_\alpha \phi \nabla^\mu\phi \nabla^\nu\nabla_\beta \phi \\\nonumber
        &~&+\kappa_9(\phi ,\rho),\\\nonumber
\rho&=&\nabla_\mu\phi \nabla^\mu\phi ,
\ea
where $\kappa_i(\phi,\rho)$, $i=1,3,8,9$ are 4 arbitrary functions of the scalar field $\phi$ and its kinetic term denoted as $\rho$ and
\ba
\label{eq:Fdef}
F_{,\rho}&=&\kappa_{1,\phi}-\kappa_3-2\rho\kappa_{3,\rho}
\ea
with $W(\phi)$ an arbitrary function of $\phi$, which means we can set it to zero without loss of generality by absorbing it  into a redefinition of $F(\phi, \rho)$.  Note that Horndeski's theory is exactly equivalent to the generalised scalar tensor theory derived by Deffayet {\it et al}, at least in four dimensions \cite{general}. This was shown explicitly in \cite{Kobayashi:2011nu}, where a useful dictionary relating the potentials in the two theories is presented. 


In his original work, Horndeski makes systematic use of the anti-symmetric Kronecker deltas which are defined by
\ba
\delta^{\mu_1...\mu_h}_{\nu_1...\nu_h}&=&\left|\begin{array}{ccc}
\delta^{\mu_1}_{\nu_1} & \hdots & \delta^{\mu_1}_{\nu_h} \\
\vdots             &        & \vdots             \\
\delta^{\mu_h}_{\nu_1} & \hdots & \delta^{\mu_h}_{\nu_h}
\end{array}\right|\\
  &=&h!\delta^{\mu_1}_{[\nu_1}...\delta^{\mu_h}_{\nu_h]}
\ea
This Lagrangian was proven to be the most general four dimensional, single-scalar tensor theory that gives second order field equations with respect to the metric $g_{\mu\nu}$ and scalar field $\phi$. Horndeski's proof is quite remarkable, not least because he starts from a very general theory of the form ${\cal L}={\cal L}(g_{\mu\nu}, g_{\mu\nu,\alpha_1},...,g_{\mu\nu,\alpha_1...\alpha_p},\phi,\phi_{,\alpha_1},...,\phi_{,\alpha_1...\alpha_q})$ with $p,q\geq 2$, thereby allowing for  higher than second derivatives in the initial Lagrangian.  Even if we neglect the scalars, this approach  is far more general than Lovelock's theorem \cite{lovelock} that initially allows only up to second derivatives of the metric field in the Lagrangian.  

The matter part of the action is given by $S_m[g_{\mu\nu}; \Psi_n]$, where we require that the matter fields are all minimally coupled to the metric $g_{\mu\nu}$.  
This follows (without further  loss of generality) from assuming that there is only violation of the {\it strong} equivalance principle and not the {\it Einstein} equivalance principle\footnote{For EEP to hold in the usual way, all matter must be minimally coupled to the {\it same} physical metric, $\tilde g_{\mu\nu}$, and this should only be a function of $g_{\mu\nu}$ and $\phi$. Dependence  on derivatives is not allowed since it would result in the  gravitational coupling to matter being momentum dependent, leading to violations of EEP.  Given $\tilde g_{\mu\nu}=\tilde g_{\mu\nu}(g_{\alpha\beta}, \phi)$, we simply compute $g_{\alpha\beta}=g_{\alpha\beta}(\tilde g_{\mu\nu}, \phi)$, and substitute back into the action (\ref{eq:action}), before dropping the tildes. Since this procedure will not generate any additional derivatives in the equations of motion, it  simply serves to  redefine the Horndeski potentials, $\kappa_i(\phi, \rho)$. }. 
Recall that this reasoning  is consistent with the original construction of  Brans-Dicke gravity~\cite{bdgravity}, where  the SEP is broken but we still impose the EEP.

The field equations emanating from this action, ${\cal E}^{\mu\nu}=-\frac{1}{\sqrt{-g} }\frac{\delta S_H}{\delta g_{\mu\nu}}, ~{\cal E}_\phi=-\frac{1}{\sqrt{-g} }\frac{\delta S_H}{\delta \phi}$, are also given by Horndeski \cite{horndeski:1974} and are of course essential in his explicit proof, relying on similar techniques to those of Lovelock \cite{lovelock}. For our purposes we will mostly make use of the Lagrange density for what follows but the equations of motions will prove crucial when we try to identify certain terms geometrically. The equations of motion obtained from (\ref{eq:horndeskiFullLag}) are ${\cal E}^{\mu\nu}=\half T^{\mu\nu}, ~ {\cal E}_\phi=0$
where  $T^{\mu\nu}=\frac{2}{\sqrt{-g}}\frac{\delta S_m}{\delta g_{\mu\nu}}$ is the energy-momentum tensor  of matter and  
\ba
\label{horndeskieom}
{\cal E}^{\epsilon}_\eta= &=& -4K_1(\phi,\rho)P^{\epsilon \alpha }{}_{\eta \mu} \nabla^\mu\nabla_\alpha\phi        -\frac{4}{3} K_{1,\rho}(\phi,\rho)\delta^{\epsilon \alpha \beta \gamma}_{\eta \mu\nu\sigma} \nabla^\mu\nabla_\alpha\phi\nabla^\nu\nabla_\beta\phi\nabla^\sigma\nabla_\gamma\phi\\\nonumber
        &~&-4P^{\epsilon \alpha }{}_{\eta \mu} K_3(\phi,\rho) \nabla_\alpha\phi\nabla^\mu\phi 
           -4 K_{3,\rho}(\phi,\rho)\delta^{\epsilon\alpha \beta \gamma}_{\eta \mu\nu\sigma} \nabla_\alpha\phi\nabla^\mu\phi\nabla^\nu\nabla_\beta\phi\nabla^\sigma\nabla_\gamma\phi\\\nonumber
        &~&-2[{\cal F}(\phi,\rho)+2{\cal W}(\phi)]G^\epsilon_\eta
           -2{\cal F}(\phi,\rho)_{,\rho}\delta_{\eta \mu\nu}^{\epsilon\alpha \beta } \nabla^\mu\nabla_\alpha\phi\nabla^\nu\nabla_\beta\phi\\\nonumber
        &~&-[2{\cal F}(\phi,\rho)_{,\phi}+4{\cal W}(\phi)_{,\phi}+\rho K_8(\phi,\rho)]\delta^{\epsilon\alpha }_{\eta \mu}  \nabla_\alpha\nabla^\mu\phi
           +K_8\delta_{\eta \mu\nu}^{\epsilon \alpha \beta } \nabla_\alpha\phi\nabla^\mu\phi\nabla^\nu\nabla_\beta\phi\\\nonumber
        &~&+K_9(\phi,\rho) \delta^{\epsilon}_{\eta}-(2{\cal F}_{,\phi \phi}+4 {\cal W}_{,\phi \phi}+\rho K_{8,\phi}+2 {K}_{9,\rho})\nabla^\epsilon \phi \nabla_\eta \phi,\nonumber
\ea
%
%
The potentials appearing here are given in terms of the action potentials by 
$$K_i=\rho \kappa_{i,\rho} \textrm{ for $i=1,3,8$}, \qquad K_9=-\half\left[\kappa_9+\rho(2(F+2W)_{, \phi \phi}+\rho \kappa_{8, \phi})\right] \qquad {\cal F}+2{\cal W}=\rho F_{, \rho}-(F+2W)$$
  Note that this expression differs slightly from the corresponding expression appearing in \cite{horndeski:1974} as we have written it in terms of the   double dual of the Riemann tensor \cite{mtw},
\be \label{def2}
P^{\mu\nu}{}_{\alpha \beta} \equiv- \frac{1}{4} \delta^{ \mu\nu \gamma \delta}_{\sigma\lambda\alpha \beta }R^{ \sigma\lambda}{}_{ \gamma \delta}=
-R^{\mu\nu}{}_{\alpha \beta} +2 R^{\mu}{}_{ [\alpha} \delta_{\beta]}^{ \nu}-2 R^{\nu}{}_{ [\alpha} \delta_{\beta]}^{ \mu}-R \delta^{\mu}_{ [\alpha} \delta_{\beta]}^{ \nu}
\ee
This object  has the same symmetry properties as the Riemann tensor, is  divergence free for all indices, and its contraction gives the Einstein tensor $P^{\mu \alpha}{}_ {\nu \alpha}=G^{\mu}_{\nu}$. It is very much analogous to the Faraday tensor in Electromagnetism. 

Because the theory is diffeomorphism invariant,  the scalar field equation of motion ${\cal E}_\phi=0$ can be derived from the following result 
\beq
\label{phimotion}
\nabla_{\mu}{\cal E}^{\mu \nu}= \half {\cal E}_\phi \nabla^{\nu} \phi 
\eeq
The important thing to note is that ${\cal E}_\phi$  is {\it still} a differential equation of second order, even though it is  a {\it derivative} of the metric equation ${\cal E}^{\mu \nu}$.

Now we want to study a cosmological setup of this theory. In other words we consider homogeneous and isotropic spatial geometries of the form,
\beq
\label{eq:cosmometric}
ds^2  = -dt^2+a^2(t)\gamma_{ij} dx^i dx^j
\eeq
where $\gamma_{ij}$  is the metric on the unit plane ($k=0$), sphere ($k=1$) or hyperboloid ($k=-1$). The following useful identities then follow,
\ba
\nabla^\mu\nabla_\nu\phi&=&diag\left(-\ddot\phi,-H\dot\phi,-H\dot\phi,-H\dot\phi\right) \label{eq:box-phi}\\
\label{eq:ricciTensor}
R^\mu_{\;\;\nu}         &=&diag\left(3\frac{\ddot a}{a},\frac{\ddot a}{a}+2H^2+2\frac{k}{a^2},\frac{\ddot a}{a}+2H^2+2\frac{k}{a^2},\frac{\ddot a}{a}+2H^2+2\frac{k}{a^2}\right)\\
\nabla^\mu\nabla_\mu\phi&=&-\ddot\phi-3H\dot\phi\\
R                       &=&6\left(\frac{\ddot a}{a}+H^2+\frac{k}{a^2}\right)\\
\label{eq:rho}
\rho&=&-\dot\phi^2
\ea
Given on the one hand, the complexity of the full action and on the other the large cosmological symmetries, we choose to initially work with the Lagrangian density rather than the equations of motion. This means that we are working within an equivalence class of Lagrangians rather than a single Lagrangian, $({\cal L},\cong)$. Any two Lagrangians are by definition within the same class, ${\cal L}\cong \tilde{{\cal L}}$ if and only if they differ by a total derivative, in particular for cosmology if they differ by a total time derivative.
In fact using (\ref{eq:cosmometric}) to (\ref{eq:rho}) above and performing several integration by parts  for each term in (\ref{eq:horndeskiFullLag}), we can arrive 
at the following rather simplified form for the cosmological minisuperspace  Lagrangian,
\ba
\label{eq:cosmolangr}
L=\frac{\int d^3 x \sqrt{-g} {\cal L}_H}{\int d^3 x \sqrt{\gamma}} \cong a^3\sum_{i=0..3}Z_iH^i
\ea
where the dependence of the $Z_i$ are as follows, ($i=0,1,2,3$),
\beq
\label{eq:a_dep_of_chi}
Z_i(\phi,\dot\phi,a)=X_{i}(\phi,\dot\phi)-Y_i(\phi,\dot\phi)\frac{k}{a^2},
\eeq
with
\ba
\label{eq:X0}
X_0&=&-\tilde{Q}_{7,\phi}\dot\phi+\kappa_9\\
\label{eq:X1}
X_1&=&-12(F+2W)_{,\phi}\dot\phi+3(Q_7\dot\phi-\tilde{Q}_7)+6\kappa_8\dot\phi^3\\
\label{eq:X2}    
X_2&=& 12F_{,\rho} \rho -12(F+2W)\\
\label{eq:X3}
X_3&=&8\kappa_{1,\rho}\dot\phi^3\\
\label{eq:Y0}
Y_0&=&\tilde Q_{1,\phi}\dot\phi+12\kappa_3\dot\phi^2-12(F+2W)\\
\label{eq:Y1}
Y_1&=&\tilde{Q}_1-Q_1\dot\phi\\\
\label{eq:Y2}
Y_2&=&0,\qquad Y_3=0\\
\label{eq:Q1}
-12\kappa_1&=&Q_1:=\frac{\del \tilde{Q}_1}{\del\dot\phi}\\
\label{eq:Q7}
6(F+2W)_{,\phi}-3\dot\phi^2\kappa_8&=&Q_7:=\frac{\del \tilde{Q}_7}{\del\dot\phi}
\ea
Here $\tilde{Q}_1$ and $\tilde{Q}_7$ are arbitrary functions of $\phi$ and $\dot{\phi}$ that, as it turns out, do not appear in the resulting equations of motion. Note the absence of higher than first derivatives in the above expressions. This is due to the properties of the Horndeski action and will be crucial for what follows. 

It is now straightforward to write down the field equations, including  a source from the matter sector in the form of a homogeneous cosmological fluid of  energy density $\rho_m$ and pressure $p$, minimally coupled to the metric:
\be
{\cal H}=-\rho_m, \qquad E_\phi=0, \qquad \dot \rho_m+3H(\rho_m+p)=0 \label{sys}
\ee 
where the Hamiltonian density and scalar equation of motion are respectively given by 
\ba
\label{hamiltonian}
{\cal H}&=&\frac{1}{a^3} \left[\frac{\partial {{L}}}{\partial {\dot{a}}} \dot{a}+\frac{\partial {{L}}}{\partial {\dot{\phi}}} \dot{\phi}-{{L}}\right]\nonumber \\
&=&\sum_{i=0..3}\left[(i-1)Z_i+Z_{i,\dot \phi}\dot\phi\right]H^i
\ea
and 
\ba
E_\phi&=&-\frac{d}{dt}\left[\frac{\del L}{\del \dot \phi}\right]+\frac{\del L}{\del \phi} \nonumber \\
&=&-\frac{d}{dt}\left[a^3\sum_{i=0..3}Z_{i,\dot\phi}H^i\right]+a^3\sum_{i=0..3}Z_{i,\phi}H^i, \label{eq:phi_eom}
\ea
This equation is linear in second derivatives, a fact that will be important later on. Indeed, in what follows it will be convenient to write it as
\be
E_\phi=\ddot\phi f(\phi,\dot\phi,a,\dot a)+g(\phi,\dot\phi,a,\dot a, \ddot a)
\ee
where the functions $f$ and $g$ are determined by equation (\ref{eq:phi_eom}). Note that the system  (\ref{sys}) includes the usual energy conservation law for the matter sector, and implies the equation of motion for the scale factor, $a$, derived directly from the minisuperspace Lagrangian:
\be
E_a=-\frac{1}{\sqrt{\gamma}} \frac{\delta S_m}{\delta a}=-3a^2 p 
\ee
where
\ba
E_a&=& -\frac{d}{dt}\left[\frac{\del L}{\del \dot a}\right]+\frac{\del L}{\del a} \nonumber \\ 
&=&-\frac{d}{dt}\left[a^3\sum_{i=1..3}iZ_ia^{-1}H^{i-1}\right]+\sum_{i=0..3}\left[a^{3-i}Z_i\right]_{,a}a^iH^i,\label{eq:a_eom}
\ea
So far everything we have said is true of the full Horndeski theory. We now specialise to the case of a self-tuning solution for this theory, and in doing so will discover a remarkable simplification leading to the theory being fully determined by just four arbitrary functions of the scalar field.

\section{self tuning in scalar-tensor theories} \label{self-tune}
We wish to identify the sector of Horndeski's theory that exhibits self-tuning, hence we first ask what it means for the relevant functions to self-tune, in a relatively model independent way.  To this end, we refer the reader to  the definition of the self-tuning filter given in the Introduction, and consider it in the context of a cosmological background in vacuo. The matter sector is expected to contribute a constant vacuum energy density, which we identify with the cosmological constant, $\langle \rho_{m}\rangle_\textrm{vac} =\rho_\Lambda$. According to our first filter  the vacuum energy should not have an impact on the spacetime curvature, so whatever the value of $\rho_\Lambda$,  we still want to have a portion of flat spacetime.
By the second filter this should remain true even when the matter sector goes through a phase-transition, changing the overall value of $\rho_\Lambda$ by a constant amount over an (effectively) infinitesimal time. In other words, we require that the abrupt change in the matter sector is completely absorbed by the scalar field leaving the geometry unchanged. Hence the scalar field tunes itself to each change in the vacuum energy $\rho_\Lambda$ and this has to be allowed independently of the time (or epoque) of transition. As we will see, these requirements place strong constraints on the theory (\ref{eq:horndeskiFullLag}).

To be consistent with the first filter, we are looking for cosmological solutions that are Ricci-flat, so (\ref{eq:ricciTensor}) tells us that
\ba
\label{flat}
H^2&=&-\frac{k}{a^2} 
\ea
where $k=0$ corresponds to a flat, and $k=-1$ a Milne slicing, of flat spacetime. For $k=1$ no flat spacetime slicing is possible. We shall also assume that the scalar $\phi(t)$ is a continuous function, but that $\dot \phi$ can be discontinuous.

We now go {\it on-shell-in-a} at the level of the field equations (\ref{sys}). This means we impose the condition (\ref{flat}) by inserting $ a=a_k(t) \equiv a_0+\sqrt{-k} t$, whilst leaving $\phi(t)$ to be determined dynamically. We find that
\ba
{\cal H}(\phi, \dot \phi, a, \dot a) &\to& {\cal H}_k(\phi, \dot \phi, a_k)\\
f (\phi, \dot \phi, a, \dot a) & \to & f_k (\phi, \dot \phi, a_k) \\
g (\phi, \dot \phi, a, \dot a, \ddot a) & \to & g_k (\phi, \dot \phi, a_k)
\ea 
Then, the \ons field equations  read
\be
{\cal H}_k(\phi, \dot \phi, a_k)=-\rho_\Lambda, \qquad \ddot \phi f_k (\phi, \dot \phi, a_k) + g_k (\phi, \dot \phi, a_k)=0  \label{ons}
\ee
where, in accordance with the second filter,  the matter sector contributes $\rho_\Lambda$ to the vacuum energy, where $\rho_\Lambda$ is a piecewise constant function of time.  Note that there is no {\it explicit} time dependence contained in ${\cal H}_k, ~f_k$ and $g_k$.

Consider the Hamiltonian constraint ${\cal H}_k=-\rho_\Lambda$, and observe that the right-hand side is discontinuous at a phase transition. Since $a_k(t)$ and $\phi(t)$ are continuous  it follows that  for the left-hand side to support this discontinuity, it must retain some non-trivial $\dot \phi$ dependence. In other words, ${\cal H}_k$ cannot be independent of $\dot \phi$.  This is our first constraint.

We now study the derivative of the Hamiltonian constraint. Since  $\rho_\Lambda$ jumps instantaneously at a phase transition, its time derivative (or equivalently, the pressure) is  delta-function localized at the transition time, $t=t_\star$. So, differentiating  the Hamiltonian constraint in (\ref{ons}) in a neighbourhood of  $t= t_\star$ we get
\ba
\sqrt{-k} \frac{\del {\cal H}_k}{\del a_k}+\dot\phi\frac{\del{\cal{H}}_k}{\del \phi}+\ddot\phi\frac{\del{\cal{H}}_k}{\del \dot\phi} &\propto&\delta(t-t_\star).
\ea
Again, since  $\phi$ is  continuous across the transition,  so it must be  $\ddot\phi$ that produces the delta-function. This is consistent with $\phi$ being continuous and $\dot\phi$ being discontinuous, with  $\ddot \phi$ providing the junction conditions for the phase transition at $t=t_\star$. 

Now consider the \ons scalar equation of motion from (\ref{ons}).  On the left hand side,   $\ddot\phi$ has a delta-function at the transition, but this is not supported on the right hand side of the equation. Thus we immediately conclude that  
\ba
f_k(\phi,\dot\phi, a_k) &=& 0,\\
g_k(\phi,\dot\phi, a_k)&=&0.
\ea
Let us focus on the first equation $f_k=0$, and consider it on either side of the transition. If $f_k=f_k(\phi, \dot \phi, a_k)$ contains non-trivial $\dot \phi$ dependence, then the left-hand side of this equation is discontinuous at the transition on account of the discontinuity in $\dot \phi$. Since this is not supported on the right-hand side we conclude that $f_k$ has no $\dot \phi$ dependence, or in other words, 
\be
f_k=f_k(\phi, a_k)
\ee
Note that this argument relies on the fact that there is no explicit time dependence contained in $f_k$  so there is nothing to  absorb the discontinuity in $\dot \phi$.

To constrain this even further, we differentiate the equation $f_k=0$ in a neighbourhood of $t= t_\star$. This yields
\be
\sqrt{-k} \frac{ \del f_k}{\del a_k} +\frac{ \del f_k}{\del \phi} \dot \phi=0
\ee  
Again, the discontinuity in $\dot \phi$ is not supported on the right-hand side, so we conclude that $\frac{ \del f_k}{\del \phi}=0 $, or equivalently, that
\be
f_k=f_k(a_k)
\ee
An identical argument implies that $g_k=g_k(a_k)$. Strictly speaking, the above arguments only hold in a neighbourhood of the transition time  $t= t_\star$. However, the transition (or transitions) can happen at any time, so we can extend our result to include {\it all} times.   Since $a_k\equiv a_0+\sqrt{-k}t $ is fixed, it now follows that the \ons scalar equations of motion $f_k=0, ~g_k=0$ contain no dynamics -- $f_k$ and $g_k$ must vanish {\it identically}. Put another way,  the scalar equation $E_\phi$ vanishes identically \ons and places no further constraints on the evolution of $\phi$. This kind of degeneracy at the level of the field equations might have been expected. We are asking our theory to admit the same solution (a patch of Minkowski) for a one parameter class of energy densities. Weinberg recognises the need for some degeneracy enroute to his no-go theorem \cite{nogo}, but his approach differs in that we have allowed $\phi=\phi(t)$.

This impacts on the \ons Lagrangian which we denote as $L_k=L_k(\phi, \dot \phi, a_k)$. Indeed the scalar equations of motion (\ref{eq:phi_eom}) are
\ba
&&-\frac{d}{dt}\left(\frac{\del L_k}{\del \dot\phi}\right)+\frac{\del L_k}{\del \phi}=0, \label{onshell1}\\
&\Rightarrow& \left[-L_{k,\dot\phi\dot\phi}\right]\ddot\phi+\left[- \sqrt{-k} L_{k,\dot\phi a_k}-\dot\phi L_{k,\dot\phi \phi}+L_{k,\phi}\right]=0\label{onshell2}\\
& \Rightarrow &f_k=-L_{k,\dot\phi\dot\phi}, \qquad g_k=- \sqrt{-k} L_{k,\dot\phi a_k}-\dot\phi L_{k,\dot\phi \phi}+L_{k,\phi}
\ea
For self tuning we now know that $f_k$ has to vanish, giving
\ba
\label{eq:selfTuneLlinearPhiDot}
L_k&=&\zeta_{k,\phi}(\phi,a_k)\dot\phi+\xi_k(\phi,a_k),
\ea
where the form of $\zeta_{k,\phi}(\phi,t)$ has been chosen for later convenience, but is still general. The vanishing of $g_k$ now yields,
\be
\xi_k=\sqrt{-k} \zeta_{k, a_k} (\phi,a_k)+ \nu_k(a_k)
\ee
At the end of the day expanding (\ref{eq:selfTuneLlinearPhiDot}), we find that the \ons Lagrangian is simply,
\ba
\label{eq:onShellLcon}
L_k &=& \dot \zeta_k+ \nu_k (a_k) \cong \nu_k (a_k) 
\ea
since the first term is a total derivative. 

We are almost done. However, we have yet to apply our third filter. This requires our self-tuning theory to admit a non-trivial cosmology. To appreciate what this means, we need to return to the scalar equation of motion before we went \ons. Recall that this equation vanishes identically when we impose the Ricci flat condition (\ref{flat}). There are two ways in which this can happen: either (i) $E_\phi=0$ is an {\it algebraic} equation in $H-\frac{\sqrt{-k}}{a}$ or (ii) $E_\phi=0$ is an {\it dynamic} equation in $H-\frac{\sqrt{-k}}{a}$. If it is the former, option (i), then we immediately see that the scalar equation of motion forces Minkowski space at all times, or else we are on a completely different branch of non-self tuning solutions. Clearly this would not pass through our third filter, so we embrace the latter, option (ii). This means the scalar equation of motion contains derivatives of $H-{\sqrt{-k}}/a$, or equivalently, that it is not independent of $\ddot a$. This is our final constraint.

To sum up then, our filters imply the following constraints:
\begin{description}
\item{\ref{self-tune}a:} the \ons  minisuperspace Lagrangian should  be  {\it  independent} of $\phi$ and $\dot \phi$, up to a total derivative.
\item{\ref{self-tune}b:} the \ons Hamiltonian density should {\it not be independent} of $\dot \phi$. 
\item{\ref{self-tune}c:} the full scalar equation of motion should {\it not be independent}  of $\ddot a$. \label{con3}
\end{description}
We are now ready to apply these directly to Horndeski's theory.

\section{applying the self-tuning filter to the Horndeski action}\label{self-tune-horndeski}

Let us return to the full minisuperspace  Lagrangian  (\ref{eq:cosmolangr}) in Horndeski's theory. We would like to push this theory through our self-tuning filter, now defined by the constraints \ref{self-tune}a to \ref{self-tune}c.  As a result, we infer the following conditions respectively
\begin{description}
\item{\ref{self-tune-horndeski}a:}  $ \sum_{i=0..3}  Z_i(a_k, \phi,\dot\phi)\left(\frac{\sqrt{-k}}{a_k}\right)^i=c(a_k)+\frac{1}{a_k^3} \frac{d\zeta }{dt}$, where $\zeta= \zeta(\phi, a_k)$  
\item{\ref{self-tune-horndeski}b:} $
\sum_{i=1..3}i Z_{i,\dot\phi}(a_k, \phi,\dot\phi) \left(\frac{\sqrt{-k}}{a_k}\right)^i\neq0.$ \label{con22}

\item{\ref{self-tune-horndeski}c:} {\it  Cannot} have $ Z_{i,\dot\phi}(a, \phi,\dot\phi)=0$ for each $i=1,2,3$ 
\end{description}
Note that condition \ref{self-tune-horndeski}a  implies that $\sum_{i=0..3}  Z_{i,\dot \phi}(a_k, \phi,\dot\phi) \left(\frac{\sqrt{-k}}{a_k}\right)^i=0$, and that this has been used to simplify  condition \ref{self-tune-horndeski}b. We also  see that condition \ref{self-tune-horndeski}b rules out $k=0$.  This is our first important result. Self-tuning is not possible within this class of scalar tensor theories for a homogeneous scalar and a spatially flat cosmology. There is, however, no obvious obstruction to self-tuning with a homogeneous scalar and a spatially hyperbolic cosmology ($k=-1$).  When this is the case, it is also easy to see  that condition \ref{self-tune-horndeski}b implies condition \ref{self-tune-horndeski}c.

Now, consider a Horndeski-like theory of the form 
\ba 
\tilde L&=& a^3 \sum_{i=0..3}\tilde Z_i(a,\phi,\dot\phi)H^i \nonumber \\
&=& a^3\left\{c(a)+\sum_{i=1..3}\tilde Z_i(a,\phi,\dot\phi)\left[H^i-\left(\frac{\sqrt{-k}}{a}\right)^i\right]\right\} \label{tildeL}
\ea
where 
\be \label{condtildeZ}
\sum_{i=1..3}i\tilde Z_{i,\dot\phi}(a, \phi, \dot\phi)\left(\frac{\sqrt{-k}}{a}\right)^i\neq0.
\ee
Such a theory will certainly squeeze through our self-tuning filter defined by the constraints \ref{self-tune-horndeski}a to \ref{self-tune-horndeski}c. In a sense, the Lagrangian $\tilde L$ is {\it sufficient} for self-tuning, but to what extent is it {\it necessary}?  Are there equivalent Horndeski-like Lagrangians, with $Z_i=\tilde Z_i +\Delta Z_i$, that admit the same set of self-tuning solutions? To establish this we need to demand that the tilded and untilded systems each have equations of motion that give the same dynamics.
In other words, 
\be \label{equiv}
{\cal H}=-\rho_m, \qquad E_\phi=0 \qquad \iff \qquad \tilde {\cal H}=-\rho_m, \qquad \tilde E_\phi=0
\ee
In general we would not be able to say much, as the statement (\ref{equiv})  does not  necessarily imply that, say,  $E_\phi\equiv \tilde E_\phi$, nor even $E_\phi\propto\tilde E_\phi$, as there could well be a non-linear relation between all the relevant equations.  Actually, owing to the special properties of the Horndeski Lagrangian in the self tuning limit, it turns out that this is not the case, and that in actual fact, we are forced to have
\be \label{=eom}
{\cal H}=\tilde {\cal H}, \qquad E_\phi=\tilde E_\phi
\ee
from which we infer the following relations
\ba
\label{Zcond}
\Delta Z_0= \dot{\phi} \frac{\mu_{,\phi}}{a^3}, \qquad \Delta Z_1=\frac{\mu_{,a}}{a^2}, \qquad \Delta Z_{2}=\Delta Z_3=0
\ea
where $\mu=\mu(a,\phi)$ is some arbitrary function.  These results are explicitly proven in appendix \ref{Delta}. Note that $a^3(\Delta Z_0+\Delta Z_1 H)=\dot \mu $, so a general self-tuning Lagrangian is equivalent to (\ref{tildeL}) up to the total derivative $\frac{d}{dt}\mu(a, \phi)$.

We are now in a position to fix the $X$'s and the $Y$'s as defined by equation (\ref{eq:a_dep_of_chi}) for the general self-tuning Lagrangian we have just derived. Restricting attention to $k \neq 0$, we show in appendix \ref{XY} that
\ba
\label{eq:X0sol}
X_0(\phi,\dot\phi)&=&V_0'(\phi)\dot\phi-\rho^{bare}_\Lambda\\
\label{eq:X1sol}
X_1(\phi,\dot\phi)&=&V_1'(\phi)\dot\phi+3V_0(\phi)\\
\label{eq:X2sol}
X_2(\phi,\dot\phi)+Y_0(\phi,\dot\phi)&=&V_2'(\phi)\dot\phi+2V_1(\phi) \\
\label{eq:X3sol}
X_3(\phi,\dot\phi)+Y_1(\phi,\dot\phi)&=&V_3'(\phi)\dot\phi+V_1(\phi)
\ea
where $V_0(\phi),V_1(\phi), V_2(\phi)$ and $V_3(\phi)$ are all arbitrary functions.  From these relations we may then evaluate the functions appearing in Horndeski's action using (\ref{eq:X0}) to (\ref{eq:Q7}) to get
\ba
\label{eq:kappa1}
\kappa_1&=&\frac{1}{8}V_3'(\phi)\left(1+\frac{1}{2}\ln|\rho|\right)+\frac{1}{4}A(\phi)\rho-\frac{1}{12}B(\phi)\\
\label{eq:kappa3}
\kappa_3&=&\frac{1}{16}V_3''(\phi)\ln|\rho|+\frac{1}{12}A'(\phi)\rho-\frac{1}{12}B'(\phi)+p(\phi)-\frac{1}{2}q(\phi)(1-\ln|\rho|)\\
\label{eq:kappa8}
\kappa_8&=&2p'(\phi)+q'(\phi)\ln|\rho|-\lambda(\phi)\\
\label{eq:kappa9}
\kappa_9&=& - \rho_\Lambda^{bare}+\frac{1}{2}V_1''(\phi)\rho+\lambda'(\phi)\rho^2\\
\label{eq:F2W}
F+2W&=&-\frac{1}{12}V_1(\phi)-p(\phi)\rho-\frac{1}{2}q(\phi)\rho\ln|\rho|
\ea
where now $V_1(\phi), ~V_3(\phi), ~A(\phi),~B(\phi), ~p(\phi), ~q(\phi)$ and $\lambda(\phi)$  are all arbitrary functions.  Again, this is shown in detail in appendix \ref{XY}.  One might wonder why it is that any dependence on $V_0$ and $V_2$ has dropped out. This is because one always has the freedom to shift $X_0$ and $Y_0$ by a total derivative  without altering the dynamics. By letting $X_0 \to X_0-\dot V_0$ and $Y_0 \to Y_0-\dot V_2$ it is easy to see that the contributions of $V_0$ and $V_2$ drop out of equations (\ref{eq:X0sol}) and (\ref{eq:X2sol}).

Having pushed Horndeski's theory through our self-tuning filter, we are led towards a subset of Horndeski's theory for which the potentials are given by these values. What is quite remarkable is that the self-tuning conditions have revealed the full dependance on the kinetic term $\rho$. Initially the Horndeski functions $\kappa_i$, $i=1,3,8,9$ were arbitrary functions of $\rho$ and $\phi$,  but now the self-tuning filter has reduced this to just seven functions of the scalar $\phi$.   However, it turns out that 
$\lambda(\phi)$, $B(\phi)$ and $p(\phi)$ all contribute total derivatives to the Lagrangian or equivelantly do not appear in the equations of motion\footnote{For example, if we only switch on $\lambda(\phi)$, we have $\kappa_8=-\lambda(\phi)$, and $\kappa_9(\phi)=\lambda'(\phi)\rho^2$, so that ${\cal L}_\lambda=-\lambda'(\phi)\rho^2+3\lambda(\phi) \rho\Box \phi-2\lambda(\phi) \delta^{\alpha\beta }_{\mu\nu} \nabla_\alpha \phi\nabla^\mu\phi \nabla^\nu \nabla_\beta \phi=\nabla_\mu(\lambda \rho \nabla^\mu \phi) \cong 0$. One can similarly show that $B$ and $p$ also contribute total derivatives to the overall Lagrangian.}. They can therefore be put to zero as physically irrelevant. 

The arbitrary constant $\rho_\Lambda^{bare}$  is nothing but the bare cosmological constant term.  Actually, the presence of this term serves as a good consistency check. The point is that any successful self tuning theory must admit an arbitrary term of this form. This is because the vacuum energy  renormalises  this  term, so if we  had been led to conclude that such a term were not present, that it should vanish,  then we would have effectively fine-tuned the bare cosmological constant against the vacuum energy. In fact, this is precisely how Weinberg's no go theorem \cite{nogo} works --- he finds that his generic ``self tuning" theory cannot admit an {\it arbitrary} term of the form $\rho_\Lambda^{bare} \sqrt{-g}$, so {\it self}-tuning is actually {\it fine}-tuning. In contrast, here we are finding that this  {\it arbitrary} cosmological constant  term  {\it is} allowed, so we have a genuinely self-tuning theory.

Finally we are left with four functions of $\phi$ for which we now seek their geometric origin. This is not clear in the Horndeski action or equations of motion due to the presence of Kronecker deltas which are useful for writing out the general Lagrangian but not physically intuitive for the filtered theory in question. Let us begin by rescaling the four remaining functions as follows
\be
q(\phi)=\frac{1}{2} V_{john}(\phi), \qquad A(\phi)=-\frac{3}{2} V_{paul}(\phi), \qquad V_1(\phi)=-6 V_{george}(\phi), \qquad V_3(\phi)=16V_{ringo}(\phi)
\ee
Further setting $\lambda(\phi)$, $B(\phi)$ and $p(\phi)$ to zero, we arrive at the following form for the Horndeski potentials
\ba
\kappa_1 &=&2V_{ringo}'(\phi)\left[1+\frac{1}{2}\ln(|\rho|)\right]-\frac{3}{8}V_{paul}(\phi)\rho \label{k1}\\
\kappa_3 &=& V_{ringo}''(\phi)\ln(|\rho|)-\frac{1}{8}V_{paul}'(\phi)\rho-\frac{1}{4}V_{john}(\phi)\left[1-\ln(|\rho|)\right] \\
\kappa_8 &=&\frac{1}{2}V_{john}'(\phi)\ln(|\rho|),\\
 \kappa_9 &=& -\rho_\Lambda^{bare}-3V_{george}''(\phi)\rho
\\
F+2W &=& \frac{1}{2}V_{george}(\phi)-\frac{1}{4} V_{john}(\phi)\rho\ln(|\rho|) \label{f2w}
\ea
We give the corresponding potentials  in the alternative form of Horndeski's theory derived by  Deffayet {\it et al} \cite{general} in  appendix \ref{app:dgsz}. Meanwhile, in appendix \ref{app:act}, we demonstrate that, after some integration by parts, these particular Horndeski potentials result in a self-tuning theory of  the form
\be
S_{Fab Four}=\int d^4 x \left[{\cal L}_{john}+ {\cal L}_{paul}+{\cal L}_{george}+{\cal L}_{ringo}\right. \\\left.-\sqrt{-g}\rho_{\Lambda}^{bare} \right]
+S_m[g_{\mu\nu}; \Psi_n] \label{selftun}
\ee
where the Lagrangians are given  by  equations (\ref{eq:john}) to (\ref{eq:ringo}). We have called this theory  {\it the Fab Four} because it is composed of four relatively simple and  elegant geometric terms, despite the fact that it originated from Horndeski's theory, which is certainly not simple, nor particularly elegant.

 To complete our analysis, let us present the cosmological equations  resulting from this theory. We find that  ${\cal H}=-\rho_m$, where the  Hamiltonian density, 
\be \label{hamff}
{\cal H}={\cal H}_{john}+{\cal H}_{paul}+{\cal H}_{george}+{\cal H}_{ringo}+\rho_\Lambda^{bare}
\ee
and
\ba
&&{\cal H}_{john}=3V_{john}(\phi)\dot\phi^2\left(3H^2+\frac{k}{a^2}\right) \nonumber\\
&&{\cal H}_{paul}=-3V_{paul}(\phi)\dot\phi^3H\left(5H^2+3\frac{k}{a^2}\right) \nonumber\\
&&{\cal H}_{george}=-6V_{george}(\phi)\left[\left(H^2+\frac{k}{a^2}\right)+H\dot\phi \frac{V'_{george}}{V_{george}}\right]\qquad \nonumber\\
&&{\cal H}_{ringo}=-24V'_{ringo}(\phi)\dot\phi H\left(H^2+\frac{k}{a^2}\right) \nonumber
\ea
Recall that one of our filters, \ref{self-tune}b, requires that the \ons Hamiltonian density should {\it not be independent} of $\dot \phi$. Plugging $H^2=-k/a^2$ into (\ref{hamff}), we immediately infer that 
\be \label{condV1}
\{V_{john}, V_{paul}, V_{george}\}  \neq \{0, 0, constant \}
\ee
This immediately rules out General Relativity which corresponds precisely to this forbidden combination. This makes sense, because as is well known, GR is {\it not} a self-tuning theory. It also rules out the possibility of a self-tuning theory supported entirely by Ringo. The point is that Ringo cannot give rise to a self-tuning theory ``without a little help from his friends", John, Paul, and George. When this is the case Ringo does have a non-trivial effect on the cosmological dynamics, but does not spoil self-tuning.

Now consider the scalar equation of motion. This is given by $E_\phi=0$, where 
\be \label{ephiff}
E_\phi={ E}_{john}+{ E}_{paul}+{ E}_{george}+{ E}_{ringo}
\ee
and 
\ba
&&{ E}_{john}= 6{d \over dt}\left[a^3V_{john}(\phi)\dot{\phi}\Delta_2\right]  - 3a^3V_{john}'(\phi)\dot\phi^2\Delta_2
 \nonumber\\
&&{E}_{paul}= -9{d \over dt}\left[a^3V_{paul}(\phi)\dot\phi^2H\Delta_2\right]  +3a^3V_{paul}'(\phi)\dot\phi^3H\Delta_2
 \nonumber\\
&&{ E}_{george}= -6{d \over dt}\left[a^3V_{george}'(\phi)\Delta_1\right]  +6a^3V_{george}''(\phi)\dot\phi \Delta_1 
+6a^3V_{george}'(\phi)\Delta_1^2  \nonumber\\
&&{ E}_{ringo} = -24 V'_{ringo}(\phi) {d \over dt}\left[a^3\left(\frac{\kappa}{a^2}\Delta_1 +\frac{1}{3}  \Delta_3 \right) \right]  \nonumber
\ea
Here we have defined the quantity 
\be
\Delta_n=H^n-\left(\frac{\sqrt{-k}}{a}\right)^n
\ee
which vanishes \ons for $n>0$. As a result, it is easy to see that  $E_\phi$ also vanishes automatically \ons, confirming what we had expected. However, we should note that the third filter, given by \ref{self-tune}c requires that the full scalar equation of motion should {\it not be independent} of $\ddot a$. This ensures that the self-tuning solution can be evolved to dynamically, and allows for a non-trivial cosmology.  From equation (\ref{ephiff}), we see that it means that 
\be \label{condV2}
\{V_{john}, V_{paul}, V_{george}, V_{ringo} \} \neq \{0, 0, constant, constant\}
\ee
This possibility has already been ruled out by the previous condition (\ref{condV1}). A detailed study of the cosmological dynamics will be presented in our companion paper \cite{fab4cos}.

The self-tuning filter we applied to the full Horndeski Lagrangian (\ref{eq:horndeskiFullLag}) is a well posed mathematical construct  with a special physical motivation. It is remarkable that it picks out such a beautifully geometric form that the Lagrangian needs to take. We will discuss some of their enchanting properties in more detail in our concluding section.

\section{The Fab Four: summary and outlook} \label{conc}
As we have seen, given some well motivated assumptions, {\it the Fab Four} represents the most general single scalar tensor  theory   capable of self-tuning. It is described by a remarkably simple and elegant action of the form,
\be
S_{Fab Four}[g_{\mu\nu}, \phi; \Psi_n]=\int d^4 x \left[{\cal L}_{john}+ {\cal L}_{paul}+{\cal L}_{george}+{\cal L}_{ringo}\right. \\\left.-\sqrt{-g}\rho_{\Lambda}^{bare} \right]
+S_m[g_{\mu\nu}; \Psi_n] \label{selftun1}
\ee
where 
  \begin{eqnarray}
\label{eq:john1}
{\cal L}_{john} &=& \sqrt{-g} V_{john}(\phi)G^{\mu\nu} \nabla_\mu\phi \nabla_\nu \phi \\
\label{eq:paul1}
{\cal L}_{paul} &=&\sqrt{-g}V_{paul}(\phi)   P^{\mu\nu\alpha \beta} \nabla_\mu \phi \nabla_\alpha \phi \nabla_\nu \nabla_\beta \phi \\
\label{eq:george1}
{\cal L}_{george} &=&\sqrt{-g}V_{george}(\phi) R \\
\label{eq:ringo1}
{\cal L}_{ringo} &=& \sqrt{-g}V_{ringo}(\phi) \hat G
\end{eqnarray}
and the matter fields, $\Psi_n$ couple only to the metric and not the scalar.  In order for self-tuning to be possible, we remind the reader that we must have 
\be \label{cond11}
\{V_{john}, V_{paul}, V_{george}\}  \neq \{0, 0, constant \}
\ee
Note that this rules out the GR limit, as  of course  it must, since that would not be a self-tuning theory. We also emphasize the presence of an arbitrary bare cosmological constant term. This serves as a good check of the validity of our analysis since any self-tuning theory {\it must} include such a term.

The cosmological field equations for an FRW universe and a homogeneous scalar were presented in equations (\ref{hamff}) and (\ref{ephiff}). For a generic choice of potentials satisfying the constraint (\ref{cond11}), a quick glance at these equations reveals that a Ricci flat universe and an explicitly time dependent  scalar is a dynamical fixed point for {\it any} vacuum energy. This remains true even as we pass through a phase transition upon which the cosmological  constant jumps by some finite amount. Strictly speaking, self-tuning is only possible in this instance when the spatial curvature is negative, and we evolve towards a Milne rather than a Minkowski geometry. However, this is really just a statement about our self-tuning ansatz and choice of coordinates. If we take our self-tuning Milne solution, we can change to hyperbolic coordinates such that the geometry is now (a portion of) Minkowski, with the scalar  rendered {\it in}homogeneous,  $\phi=\phi(|x|^2-t^2)$.

Beyond cosmology, the full  {\it Fab Four} equations of motion are given by
\ba
&& {\cal E}^{\mu \nu}_{john}+{\cal E}^{\mu \nu}_{paul}+{\cal E}^{\mu \nu}_{george}+{\cal E}^{\mu \nu}_{ringo}=\half T^{\mu \nu}  \\
&& {\cal E}^\phi_{john}+{\cal E}^\phi_{paul}+{\cal E}^\phi_{george}+{\cal E}^\phi_{ringo}=0
\ea
where the contribution of each term from variation of the metric is given by 
\ba
&&{\cal E}^{\eta \epsilon}_{john} = V_{john}(\rho G^{\eta \epsilon}-2 P^{\eta \mu \epsilon \nu}  \nabla_\mu \phi \nabla_\nu \phi  )+\half g^{\epsilon \theta}\delta^{\eta \alpha \beta}_{\theta \mu\nu} \nabla ^\mu  (\sqrt{V_{john}} \nabla_\alpha \phi) \nabla^\nu ( \sqrt{V_{john}} \nabla_\beta \phi ) \\
&& {\cal E}^{\eta \epsilon}_{paul} =\frac{3}{2} P^{\eta \mu \epsilon \nu} \rho V_{paul}^{2/3}  \nabla_{\mu}  \left(V_{paul}^{1/3}  \nabla_\nu \phi \right)  
+\half g^{\epsilon \theta} \delta^{\eta \alpha \beta \gamma}_{\theta \mu\nu\sigma}  \nabla^\mu \left(V_{paul}^{1/3} \nabla_\alpha\phi\right) \nabla^\nu  \left(V_{paul}^{1/3} \nabla_\beta\phi\right) \nabla^\sigma \left(V_{paul}^{1/3} \nabla_\gamma\phi\right) \\
&& {\cal E}_{george}^{\eta \epsilon}= V_{george} G^{\eta \epsilon}-(\nabla^\eta \nabla^\epsilon - g^{\eta \epsilon} \Box )V_{george} \\
&& {\cal E}^{\eta \epsilon}_{ringo}=-4 P^{\eta \mu \epsilon \nu} \nabla_\mu \nabla_\nu V_{ringo}
\ea
and from variation of the scalar by 
\ba
&&{\cal E}^\phi_{john} = 2 \sqrt{V_{john}} \nabla_\mu (\sqrt{V_{john}} \nabla_\nu \phi) G^{\mu\nu} \\
&&{\cal E}^\phi_{paul} = 3 V_{paul}^{1/3}  \nabla_\mu \left(V_{paul}^{1/3} \nabla_\alpha\phi\right) \nabla_\nu  \left(V_{paul}^{1/3} \nabla_\beta\phi\right)P^{\mu\nu \alpha \beta} -\frac{3}{8} V_{paul} \rho \GB \\
&&{\cal E}^\phi_{george} = -V_{george}' R \\
&&{\cal E}^\phi_{ringo} = -V_{ringo}' \GB
\ea
Note that we have absorbed $\rho_\Lambda^{bare}$ into a renormalisation of  the energy momentum tensor $T^{\mu \nu}$. Again, we emphasize the fact that the scalar equation of motion vanishes trivially on (a portion of) Minkowski space. 

{\it The Fab Four} should generally be considered in combination, and not as individuals.   We have already seen how the constraint  (\ref{cond11}) suggests that Ringo should not be considered in isolation. The point is that  on a would be self-tuning solution, the geometry is  Minkowski space and so  ${\cal E}^{\mu \nu}_{ringo} \to 0$. This means that  Ringo in isolation cannot support a non-vanishing vacuum energy and so self-tuning is destroyed.  George is another term that should not be considered in isolation, but for more phenomenological reasons. This is because  it corresponds to Brans-Dicke gravity with Brans-Dicke parameter $w=0$. Such a theory would never pass solar system gravity tests for which one typically needs $w>40000$. 

It is natural to wonder whether or not there is a phenomenologically viable version of {\it the Fab Four}. The case of George in isolation might give us cause for concern. Indeed, whatever {\it Fab Four} terms we include it is clear that our theory contains a light scalar that is giving rise to a considerable modification of General Relativity. Is it possible to suppress this modification at the relevant scales in order to pass solar system constraints? To this end, we are cautiously optimistic as we will now explain. We see that George already contains a GR like contribution if we write its potential as
$$V_{george}=\frac{1}{16 \pi G_N} + \Delta V_{george}$$
Thus a general {\it Fab Four} theory can be written as $S_{FabFour}=S_{GR}+\Delta S$, where $S_{GR}$ is the action for General Relatvity, and $\Delta S$ encodes the modification, including contributions from the potentially troublesome light scalar. However, we now note that John and Paul contain non-trivial derivative interactions  and if they are present in $\Delta S$, then we have all the necessary ingredients in order to invoke the Vainshtein mechanism \cite{vainsh}. This is a process by which an additional light degree of freedom is screened at short distances around a heavy source. It was originally studied in the context of massive gravity \cite{vainsh}  but has since been widely explored in DGP gravity \cite{dgpvainsh} and galileon theories \cite{galileon}. The presence of derivative interactions of the additional mode causes linearised perturbation theory to break down at larger than expected scales -- the Vainshtein scale. Below the Vainshtein scale  the field lines associated with the additional mode are diluted and one is able to recover GR to good approximation \cite{hirsute}. The Vainshtein scale depends on the mass of the source, so typically for the Sun one would like this to exceed the size of the solar system. For these reasons we expect any phenomenologically viable theory of {\it the Fab Four} to contain at least one of either John or Paul.  Vainshtein effects in some subclasses of Horndeski's theory have been studied recently \cite{vainshgal}.

We also need {\it the Fab Four} to recover a sensible cosmological evolution. Vainshtein effects are typically absent in background cosmology owing to the large amount of symmetry, so we cannot appeal to the above arguments in this instance. However, in our companion paper we have been able to show explicitly that sensible cosmological solutions are possible \cite{fab4cos}. Here one assumes a large vacuum energy that completely dominates the energy density of the Universe. For certain choices of potential we can show that this vacuum energy can actually mimic a matter dominated expansion.   On the subject of cosmology, it is worth noting that  recently John has   been used in some models of Higgs inflation \cite{higgs}, whilst  John, Paul and George have been used as a proxy theory for studying cosmological solutions of massive gravity \cite{cosmogal}.

Given an interesting solution to a {\it Fab Four} theory (ie. one that has a sensible cosmology and passes solar system tests),  we need to check if it is perturbatively stable. In particular, does the spectrum of perturbations contain ghost or gradient instabilities, and if so, how bad are they? It is difficult to make any generic statements, mainly because the spectrum of  solutions is potentially so vast given the fact that we have four arbitrary potentials.  What we can say is that instabilities are not necessarily automatic in {\it the Fab Four}. Although not phenomenologically viable, the case of Brans-Dicke gravity with $w=0$ discussed earlier is certainly free of ghosts and tachyons. Perhaps the most sensible approach is   to find the phenomenologically viable solutions first, and then test their stability.

Of course,  the classical {\it Fab Four} Lagrangian will inevitably receive radiative corrections from matter and/or gravity loops. If these corrections are large then it is clear that the classical self-tuning solutions should not be trusted.  Again, this is a difficult question to address properly without a better understanding of the preferred background solutions, and preferred potentials. The reason is that such corrections are sensitive to the cut-off which itself is sensitive to the background, which in turn is sensitive to the potentials. Therefore a detailed analysis of this should probably be postponed until after we have exhausted  other issues such as cosmology, solar system tests, and stability.  In other words we first obtain a class of sensible cosmological solutions and potentials and investigate the radiative corrections about these in detail. Having said that, an heuristic analysis of radiative corrections about the self-tuning vacuum solution reveals that it might well be possible to render some {\it  Fab Four} theories  safe from large quantum corrections. This is discussed in detail in appendix \ref{radiative}.  There we show that  radiative corrections on the self-tuning  background can be suppressed provided the  cut-off of the effective theory  $\Lambda_{UV}$ satisfies the inequality 
$$\sqrt{{G}_\textrm{eff}\rho_\Lambda}<\Lambda_{UV} < \rho_\Lambda^{1/4}$$
 where ${G}_\textrm{eff}$ is the (possibly time dependent) strength of the gravitational coupling to matter, in the {\it  linearised} regime. Typically we might expect $\rho_\Lambda^{1/4} \sim$ TeV and ${G}_\textrm{eff} \sim M_{pl}^{-2}$, so this condition is far from restrictive.  Note that a more detailed analysis of radiative corrections might well be sensitive to the elegant geometrical  structure  of {\it the Fab Four} terms,

Let us now discuss that elegant structure. The first thing to note is that each member of {\it the Fab Four}  vanishes for vanishing curvature. This stems from the self-tuning nature of the theory. As we saw from the scalar equations of motion, each term imposes a constraint that  is satisfied automatically in Minkowski space.  Another feature of {\it the Fab Four} terms is that they only give rise to second order field equations. This had to be the case, of course, since they represent a special case of Horndeski's theory. We also note that each of {\it the Fab Four} appear in the Kaluza-Klein reduction of Lovelock theory \cite{kkl}, from which  they inherit the second order equations of motion. 
This is obvious for John, George and Ringo  \cite{kkl}  but also turns out to be true of  Paul  which 
originates from the third order Lovelock curvature invariant \cite{VanAcoleyen:2011mj}.

It is instructive to see how exactly second order field equations are achieved given the form of each individual member of {\it the Fab Four}. For George and Ringo, the presence of the Euler Densities, $\sqrt{-g} R$ and $\sqrt{-g} \GB$ are crucial in this respect. Indeed, both terms take the form
$$ V(\phi) (\textrm{Euler density}) $$
These are the only possibilities of the form $\sqrt{-g} V(\phi) {\cal Q}$, where $\cal Q$ is a  non-trivial scalar constructed out of the curvature, because any other choice would have led to higher order field equations. 

For John and Paul, the fact that there are curvature terms contracted with derivatives of the scalar is potentially worrying, since generically this would also lead to higher order field equations. However the key point is that both terms take the form
$$V(\phi) \nabla_\mu \phi \nabla_\nu \phi \frac{\delta W}{\delta g_{\mu\nu}}$$
where $W=W[g_{\mu\nu}, \phi]$ is some diffeomorphism invariant superpotential, with second order Euler-Lagrange equations.  The diffeomorphism invariance of $W$ ensures that $\del_\mu \left(\frac{\delta W}{\delta g_{\mu\nu}} \right)\equiv 0$, and this helps to  protect us from developing higher order terms in the equations of motion. The superpotentials themselves are given by
\be
W_{john}=-\int d^4 x \sqrt{-g} R, \qquad W_{paul}=\frac{1}{4} \int d^4 x \sqrt{-g} \phi  \GB  
\ee
Here we see the Euler densities appearing again.  In fact, we can go a little further and identify a certain hierarchy within the structure of {\it the Fab Four}. In particular, we note that John's superpotential is a George type term, and that Paul's superpotential is a Ringo type term. In other words, John is a derivative of George whilst Paul is a derivative of Ringo.  This geometric structure certainly lends itself to generalising {\it the Fab Four} to multiple scalar fields.

We end our discussion by emphasizing the true purpose of this work. Rather than presenting a solution to  the cosmological constant problem, we are more interested learning about the nature of the problem and the tools you might need to tackle it. In this respect our work is in the same spirit as Weinberg's no-go theorem \cite{nogo}.  Through this theorem, Weinberg presented  a carefully chosen set-up, and then discovered that one was inevitably  faced with an inpenetrable barrier to solving the problem. By relaxing the condition of Poincar\'e invariance at the level of the self-adjusting fields, we have changed the rules of the game slightly. We have used Horndeski's very general theory as the arena in which we intend to study the problem, and having changed the rules, we have been able to pass through  Weinberg's barrier. Of course, only a tiny fraction of Horndeski's theory made it through. This is {\it the Fab Four}. How much further can they go? Clearly there are a number of extra barriers to overcome, including solar system tests, cosmological tests, and questions about stability and naturalness, as we have just discussed.  Each of these barriers will reduce the size of the arena by ruling out certain choices of {\it Fab Four} potentials and the corresponding solutions. Will there be anything left once we have taken on {\it all} of the barriers? This is impossible to say at this early stage, but one thing we can say is that whatever happens we will learn something important about the cosmological constant problem and how to tackle it. Should {\it the Fab Four} ultimately fail in tackling $\Lambda$, then we will essentially have a new no-go theorem.  This is because our starting point was a very general class of models -- all second order scalar tensor theories -- so {\it the Fab Four}'s  failure would also be the failure of all theories within this very general class. As with Weinberg's theorem, we could then ask how exactly this failure came about,  in the hope that it might point towards new directions and new approaches.  The other possibility, of course, is that some particular {\it Fab Four} Lagrangians do make it through every barrier, in which case we are left with an extremely interesting resolution of the cosmological constant problem.

\begin{acknowledgments} 
EJC and AP acknowledge  financial support from the Royal Society and CC from STR-COSMO, ANR-09-BLAN-0157. 
\end{acknowledgments}

\appendix
\section{Proof that ${\cal H}=\tilde {\cal H}$ and $E_\phi=\tilde E_\phi$, and calculation of $\Delta Z_i$} \label{Delta}
Our starting point is two Horndeski theories, defined by (\ref{eq:cosmolangr}) and (\ref{tildeL}), satisfying the criteria for equivalence given by (\ref{equiv}). We begin with the Hamiltonian constraints. In principle these differ by a function $\Delta {\cal H}=\Delta {\cal H}(a, \dot a, \phi, \dot \phi)$, as follows
\be
{\cal H}+\rho_m \equiv  \tilde {\cal H}+\rho_m+ \Delta {\cal H}
\ee
The functional dependence of $\Delta {\cal H}$ is on account of the fact that matter couples in the same way in both our theories (by assumption). From (\ref{equiv}) we require that $\Delta {\cal H}$ should  vanish on-shell whenever $\tilde {\cal H}=-\rho_m, ~\tilde E_\phi=0$. However, since  $\Delta {\cal H}$ is independent of $\rho_m$ it cannot vanish by virtue of the equation $\tilde {\cal H}=-\rho_m$. Similarly, since it is independent of $\ddot a$, nor can it vanish by virtue of $\tilde E_\phi=0$, which is necessarily dependent on $\ddot a$ by condition \ref{self-tune}c above. If $\Delta {\cal H}$ does not vanish by virtue of $\tilde {\cal H}=-\rho_m$ or $\tilde E_\phi=0$ we must conclude that it vanishes identically. In other words
\ba
\label{eq:equateHamsFinal}
{\cal H} \equiv \tilde {\cal H}.
\ea
This is a rather strong constraint with useful implications. Given that  $\Delta Z_i=Z_i-\tilde Z_i$ we see that it implies
\be
\Delta {\cal H}=\sum_{i=0..3}\left[(i-1)\Delta Z_i+\Delta Z_{i,\dot \phi}\dot\phi\right]H^i \equiv 0
\ee
Equating powers of $H$ gives
\ba
\label{eq:deltaZcondHam}
(i-1)\Delta Z_i+\Delta Z_{i,\dot\phi}\dot\phi& \equiv &0 \qquad i=0 \ldots 3.
\ea
and, so we integrate to find that
\be \label{DeltaZ}
\Delta Z_i=\sigma_i(a, \phi) \dot \phi^{1-i}
\ee
We now turn our attention to the scalar equation of motion. These differ by a function $\Delta E_\phi=\Delta E_\phi(a, \dot a, \ddot a \phi, \dot \phi, \ddot \phi)$, as follows
\be
E_\phi \equiv \tilde E_\phi+\Delta E_\phi
\ee
As above, since  $\Delta E_\phi$ is independent of $\rho_m$ it cannot vanish by virtue of the equation $\tilde {\cal H}=-\rho_m$.  At best it vanishes by virtue of the equation  $\tilde E_\phi=0$. To proceed a little further we note that equation (\ref{eq:phi_eom}) suggests that $E_\phi$ can be written in the form
\be
E_\phi=\ddot a \alpha+\ddot \phi \beta+ \gamma
\ee
where
\ba
\alpha(a, \dot a,  \phi, \dot \phi) &=&-a^2\sum_{i=0..3} i Z_{i, \dot \phi}  H^{i-1} \\
\beta (a, \dot a, \phi, \dot \phi)&=& -a^3 \sum_{i=0..3} Z_{i, \dot \phi \dot \phi} H^i \\
\gamma(a, \dot a, \phi, \dot \phi)  &=& -a^3 \sum_{i=0..3} \left[\left((i+3) Z_{i,  \dot \phi} +a Z_{i, \dot \phi a}\right)H+\dot \phi Z_{i, \phi \dot \phi}-Z_{i, \phi}\right] H^i
\ea  
with  similar expressions for $\tilde E_\phi$, $\tilde \alpha, \tilde \beta$ and $\tilde \gamma$, and by association, for $\Delta E_\phi$, $\Delta\alpha, \Delta\beta$ and $\Delta\gamma$. Now, since 
\be
\ddot a=\frac{1}{\tilde \alpha}(\tilde E_\phi -\ddot \phi \tilde \beta- \tilde \gamma)
\ee
we see that we can write
\be
\Delta E_\phi =\frac{\Delta \alpha}{\tilde \alpha} \tilde E_\phi +\ddot \phi \frac{\tilde \alpha \Delta \beta-\tilde \beta \Delta \alpha}{\tilde \alpha} + \frac{\tilde \alpha \Delta \gamma-\tilde \gamma \Delta \alpha}{\tilde \alpha}
\ee
Note that $\tilde \alpha \neq 0$ on account of condition \ref{self-tune}c. Because $\Delta E_\phi$ ought to vanish by virtue of $\tilde E_\phi=0$, we immediately infer that
\be \label{condEab}
\Delta E_\phi =\frac{\Delta \alpha}{\tilde \alpha} \tilde E_\phi,  \qquad \tilde \alpha \Delta \beta=\tilde \beta \Delta \alpha, \qquad \tilde \alpha \Delta \gamma=\tilde \gamma \Delta \alpha
\ee
However, we know from equation (\ref{DeltaZ}) that
\ba
\Delta \alpha &=& -a^2\sum_{i=0..3} i (1-i)\sigma_i \frac{H^{i-1}}{\dot \phi^i} \label{Deltaal}\\
\Delta \beta &=& -a^3 \sum_{i=0..3} i (i-1)\sigma_i \frac{H^{i}}{\dot \phi^{i+1}} \\
\Delta \gamma &=&  -a^3 \sum_{i=0..3} \left[ ((i+3) \sigma_i+a\sigma_{i, a} )H(1-i)-i \sigma_{i, \phi} \dot \phi\right] \frac{H^i}{\dot \phi^i} \label{Deltaga}
\ea 
It follows from the condition $ \tilde \alpha \Delta \beta=\tilde \beta \Delta \alpha$ that {\it  unless $\Delta E_\phi$ vanishes identically}, we must have 
\begin{align}
&&aH \tilde \alpha &=-\dot \phi \tilde \beta&~ \nonumber \\
&\implies & \sum_{i=0..3} i \tilde Z_{i, \dot \phi}  H^{i} &=-\sum_{i=0..3}  \tilde Z_{i, \dot \phi \dot \phi}\dot \phi  H^i &\nonumber \\
&\implies   &i \tilde Z_{i, \dot \phi}  &=- \tilde Z_{i, \dot \phi \dot \phi}\dot \phi & \nonumber \\
&\implies & \tilde Z_i &=u_i(a, \phi){\cal I}_i(\dot\phi)+v_i(a, \phi) & \label{Zi}
\end{align}
where ${\cal I}_i(\dot\phi)=\begin{cases}\dot \phi^{1-i}& \textrm{for $i \neq 1$}\\\ln  \dot \phi& \textrm{for $i = 1$}  \end{cases}$. Now from equation (\ref{Zi}) and  the definition of $\tilde L$ given by equation (\ref{tildeL}),  we have that
\be
c(a) =\sum_{i=0..3} \tilde Z_i \left(\frac{\sqrt{-k}}{a}\right)^i=\sum_{i=0..3} (u_i(a, \phi){\cal I}_i(\dot\phi)+v_i(a, \phi)) \left(\frac{\sqrt{-k}}{a}\right)^i
\ee
Equating powers of $\dot \phi$, we see that $u_i=0$ for all $i$, and so it immediately follows that $\tilde Z_{i, \dot \phi}=0$ for all $i$, which contradicts the condition (\ref{condtildeZ}). We are therefore forced to accept the alternative possibility  that $\Delta E_\phi$ vanishes identically. Thus we have proven equation (\ref{=eom}).

It remains to prove (\ref{Zcond}). We now know that $\Delta \alpha \equiv 0$, where $\Delta \alpha$ is given by (\ref{Deltaal}). Equating powers of $H$ we immediately see that $\sigma_2\equiv \sigma_3 \equiv 0$.  Furthermore, $\Delta \gamma \equiv 0$ where $\Delta \gamma$ is given by (\ref{Deltaga}), yielding the relation
\be
\sigma_{1, \phi}=3 \sigma_0+a\sigma_{0, a} \qquad \implies \qquad a^3 \sigma_0=\mu_{, \phi}, \qquad a^2\sigma_1=\mu_{, a}
\ee
where $\mu=\mu(a, \phi)$. Equation (\ref{Zcond}) follows automatically.

\section{Derivation of the Horndeski potentials in the self-tuning theory.} \label{XY}
Having identified the general form for the minisuperspace Lagrangian for the self-tuning Horndeski theory, we would like to derive the form of the corresponding Horndeski potentials. To this end, we first need to calculate the $X$'s and the $Y$'s as defined by equation (\ref{eq:a_dep_of_chi}).  Comparing this with the general form of the self-tuning Lagrangian, $L=\tilde L+\frac{d}{dt} \mu(a, \phi)$, where $\tilde L$ is given by equation (\ref{tildeL}), we find that
\ba
c(a)-\sum_{i=1..3}\tilde Z_i\left(\frac{\sqrt{-k}}{a}\right)^i+a^{-3}\dot\phi \mu_{,\phi}&=&X_0(\phi,\dot\phi)-\frac{k}{a^{2}}Y_0(\phi,\dot\phi) \label{Z0}\\
\tilde Z_1+a^{-2}\mu_{,a}&=&X_1(\phi,\dot\phi)-\frac{k}{a^{2}}Y_1(\phi,\dot\phi) \label{Z1}\\
\tilde Z_i &=&  X_i(\phi, \dot \phi), \qquad i=2, 3 \label{Z23}
\ea
Substituting (\ref{Z1}) and (\ref{Z23}) into (\ref{Z0}) gives the relation,  
\be
\label{cofa}
c(a)-\frac{\sqrt{-k}}{a}\left[X_1-\frac{k}{a^{2}}Y_1-a^{-2}\mu_{,a}\right]
      -\sum_{i=2,3}X_i\left(\frac{\sqrt{-k}}{a}\right)^i+a^{-3}\mu_{,\phi}\dot\phi 
     =X_0(\phi,\dot\phi)-\frac{k}{a^{2}}Y_0(\phi,\dot\phi)
\ee
We now restrict attention to $k \neq 0$,  and solve this equation by expanding $c$ and $\mu$ as  power series in $\sqrt{-k}/a$
\be
c(a)=\sum_{i=-\infty..\infty}c_i\left(\frac{\sqrt{-k}}{a}\right)^i, \qquad 
a^{-3}\mu =\sum_{i=-\infty..\infty}h_i(\phi)\left(\frac{\sqrt{-k}}{a}\right)^i
\ee
Plugging this into (\ref{cofa}), and  equating powers of $\sqrt{-k}/a$, we find that
\ba
X_0 &=& c_0 +\dot h_0+4 h_{-1} \label{X0}\\
X_1&=& c_1+\dot h_1+3 h_0 \\
X_2+Y_0 &=&c_2+\dot h_2+2h_1 \\
X_3+Y_1 &=&  c_3+\dot h_3+h_2 \label{X3}
\ea
along with the relation  
\be
c_i+\dot h_i+(4-i)h_{i-1}=0\qquad i\leq -1\textrm{~or~} i \geq 4
 \ee
This last equation  is readily solved by defining
\ba
V_i&=&h_i+\frac{c_{i+1}}{3-i}\qquad i\neq3, \qquad V_3=h_3
\ea
so that we have
\ba
V'_i(\phi) \dot\phi+(4-i)V_{i-1}&=&0\qquad i \leq -1\textrm{~or~} i \geq 4
\ea
Since $V_i$ does not depend on $\dot\phi$ it follows that
\ba
V_{-1}&=&const,\;V_{-2}=V_{-3}=...=0,\;V_4=V_5=...=0
\ea
Plugging everything back into equations (\ref{X0}) to (\ref{X3}) we obtain
\ba
X_0&=&V_0'\dot\phi+4V_{-1}=4(const)+V_0'\dot\phi \label{X0A}\\
X_1&=&V_1'\dot\phi+3V_0 \label{X1A} \\
X_2+Y_0&=&V_2'\dot\phi+2V_1\\
X_3+Y_1&=&V_3'\dot\phi+V_2
\ea
Identifying $const=-\frac{1}{4} \rho_\Lambda^{bare}$, we arrive at equations (\ref{eq:X0sol}) to (\ref{eq:X3sol}). 

To calculate the precise form of the Horndeski potentials, we make use of the basic relations (\ref{eq:X0}) to  (\ref{eq:Q7}), (\ref{eq:Fdef}) and (\ref{eq:rho}) along with our newly derived formulae (\ref{eq:X0sol}) to (\ref{eq:X3sol}). We  shall begin by deriving $\kappa_9$.  First combine (\ref{eq:X1}) and (\ref{eq:Q7}) to get the relation  
\be
X_1 =\tilde Q_{7,\phi}\dot\phi-3\tilde Q_7=\dot\phi^4(\tilde Q_7/\dot\phi^3)_{,\dot\phi} \label{X1AB}
\ee
Using equation (\ref{eq:X1sol}), one can straightforwardly  integrate  (\ref{X1AB}) to obtain
\ba \label{Q7}
\tilde Q_7&=&-V_0-\frac{1}{2}V_1'\dot\phi+\lambda(\phi)\dot\phi^3
\ea
where $\lambda(\phi)$ is an arbitrary function of integration. Given that $\rho=-\dot \phi^2$, we can use this result, along with equations (\ref{eq:X0sol}) and (\ref{eq:X0}) to derive the formula (\ref{eq:kappa9})  for $\kappa_9$.

Next we derive $\kappa_1$. From  (\ref{eq:X3}), and (\ref{eq:Y1}) we have that
\ba
X_3+Y_1 =8\kappa_{1,\rho}\dot\phi^3-\tilde Q_{1,\dot\phi}\dot\phi+\tilde Q_1=\frac{\dot \phi^4}{3} \left[(\tilde Q_1/\dot\phi)_{,\dot\phi}/\dot\phi \right]_{,\dot\phi}
\ea
where  in the second relation  we have used (\ref{eq:Q1}) and the fact that $\del_\rho=-\frac{1}{2\dot \phi} \del_{\dot \phi}$. Using equation (\ref{eq:X3sol}), this yields
\ba \label{Q1}
\tilde Q_1&=&V_2-\frac{3}{2}\dot\phi V_3'\ln\dot\phi+A(\phi)\dot \phi^3+B(\phi)\dot\phi
\ea
 where $A(\phi)$ and $B(\phi)$ are arbitrary functions of integration.  We then use $\kappa_1=-\frac{1}{12} \tilde Q_{1, \dot \phi}$ and $\rho=-\dot \phi^2$, to arrive at equation (\ref{eq:kappa1}).

We shall now derive $F+2W$.  From (\ref{eq:X2}), (\ref{eq:Y0}), and  (\ref{eq:Q1}), we have that
\be
X_2+Y_0=\dot \phi \tilde Q_{1, \phi}-12\dot \phi^2 F_{, \rho}-24(F+2W)+12 \dot \phi^2 \kappa_3
\ee
and using equation (\ref{eq:X2sol}) we obtain 
\ba
\label{eq:clm7Intermediate}
\tilde Q_{1,\phi}+12\left[\kappa_3\dot\phi-F_{,\rho}\dot\phi-2(F+2W)/\dot\phi\right]=\frac{2V_1}{\dot\phi}+V_2'
\ea
Differentiating this with respect to $\dot\phi$, and making use (\ref{eq:Q1}) and (\ref{eq:Fdef}) we arrive at the following differential equation for $F+2W$,
\ba
-\frac{V_1}{12}&=&\rho^2 F_{,\rho\rho}-\rho F_{,\rho}+(F+2W)
\ea
This is easily integrated to give the formula (\ref{eq:F2W}) for $F+2W$, where $p(\phi)$ and $q(\phi)$ are arbitrary functions of integration.

Moving on to $\kappa_3$. The formula  (\ref{eq:kappa3}) now follows immediately from equation (\ref{eq:clm7Intermediate}), once we plug in our solutions (\ref{eq:F2W}) and (\ref{Q1}) for $F+2W$ and $\tilde Q_1$ respectively. Similarly the solution for $\kappa_8$ given by (\ref{eq:kappa8}) also follows immediately from the solutions (\ref{eq:F2W}) and (\ref{Q7}) for $F+2W$ and $\tilde Q_7$ respectively.

\section{DGSZ potentials for   {\it the Fab Four}} \label{app:dgsz}
It was shown in \cite{Kobayashi:2011nu} that in four dimensions Horndeski's theory is equivalent to the generalised galileon theory derived independently by Deffayet {\it et al} \cite{general}. This latter theory is given by the Lagrangian density
\begin{multline}
{\cal L}_{DGSZ}=K(\phi, X)-G_3(\phi, X) \Box \phi+G_4 (\phi, X)R+G_{4, X} \left[ (\Box \phi)^2-(\nabla_\mu \nabla_\nu \phi)^2 \right]
\\
+G_5(\phi, X) G_{\mu\nu}\nabla^\mu \nabla^\nu \phi-\frac{G_{5, X}}{6} \left[(\Box \phi)^3-3\Box \phi (\nabla_\mu \nabla_\nu \phi)^2 +2(\nabla_\mu \nabla_\nu \phi)^3 \right]
 \end{multline}
and $X=-\half \nabla_\mu \phi \nabla^\mu \phi=-\half \rho$. The dictionary relating the potentials in the two theories is also presented in \cite{Kobayashi:2011nu}, 
\ba
K&=& \kappa_9+\rho \int^\rho d\rho' \left(\kappa_{8, \phi}-2\kappa_{3, \phi\phi}\right)\\
G_3 &=& 6(F+2W)_{, \phi}+\rho \kappa_8+4\rho  \kappa_{3, \phi}-\int^\rho d\rho' \left(\kappa_{8}-2\kappa_{3, \phi}\right) \\
G_4 &=& 2(F+2W)+2\rho \kappa_3 \\
G_5 &=& -4\kappa_1
\ea
Substituting (\ref{k1}) to (\ref{f2w})  into these formulae, and neglecting terms that contribute an overall total derivative, we obtain the following DGSZ potentials for {\it the Fab Four}, 
\ba
K&=&-\rho_\Lambda^{bare}+ 2V_{john}''(\phi)X^2-V_{paul}'''(\phi)X^3+6V_{george} ''(\phi) X+8V_{ringo}''''(\phi)X^2(3- \ln(|X|))  \\
G_3 &=&3 V_{john}'(\phi) X-\frac{5}{2} V_{paul}''(\phi)X^2+3 V_{george}'(\phi)+ 4V_{ringo}''' X(7-3\ln(|X|))  \\
G_4 &=&  V_{john}(\phi) X-V_{paul}'(\phi) X^2 +V_{george}(\phi)+4V_{ringo}''(\phi) X(2- \ln (|X|)) \\
G_5  &=&-3V_{paul}(\phi)X-4V_{ringo}'(\phi)\ln(|X|)
\ea
\section{From Horndeski's potentials to {\it the Fab Four}: metric equations of motion} \label{app:act}
We now show how the Horndeski potentials for  {\it the Fab Four} do indeed give rise to a theory of the form (\ref{selftun}). To this end, it is sufficient to show the equivalence of the equations of motion. We begin with John's contribution. The non-zero Horndeski potentials are
\ba
\kappa_3&=&-\frac{1}{4}V_{john}(\phi)(1-\ln|\rho|)\\
\kappa_8&=&\half V'_{john}(\phi)\ln|\rho|\\
F+2W&=&-\frac{1}{4}V_{john}(\phi)\rho\ln|\rho|
\ea
which translate to the following non-zero potentials appearing in the equations of motion,
\beq
K_3=\frac{1}{4}V_{john}, \qquad K_8=\half V'_{john}, \qquad {\cal F}+2{\cal W}=-\frac{1}{4}V_{john} \rho
\eeq
Using the expression (\ref{horndeskieom}), we see that
\be
{\cal E}^{\eta \epsilon }_{john}=\half V_{john}(\rho G^{\eta \epsilon }-2 P^{\eta \mu \epsilon \nu}  \nabla_\mu \phi \nabla_\nu \phi  )+\half g^{\epsilon \theta}\delta^{\eta \alpha \beta}_{\theta \mu\nu} \nabla^\nu \nabla_\beta \phi(V_{john} \nabla ^\mu  \nabla_\alpha \phi +V_{john}' \nabla^\mu \phi \nabla_\alpha \phi) 
\ee
After the tedious expansion of the  final Kronecker delta the equations of motion are recognised as those derived upon varying  $\int d^4 x {\cal L}_{john}$, where ${\cal L}_{john}= \sqrt{-g} V_{john}(\phi) G_{\mu\nu} \nabla^\mu \phi \nabla^\nu \phi$ (see, for example, \cite{CDD}). Note that this equation can be more succintly written as
\ba
{\cal E}^{\eta \epsilon }_{john} &=& V_{john}(\rho G^{\eta \epsilon }-2 P^{\eta \mu \epsilon \nu}  \nabla_\mu \phi \nabla_\nu \phi  )+\half g^{\epsilon \theta}\delta^{\eta \alpha \beta}_{\theta \mu\nu} \nabla ^\mu  (\sqrt{V_{john}} \nabla_\alpha \phi) \nabla^\nu ( \sqrt{V_{john}} \nabla_\beta \phi ) \ea
We now turn to Paul. The non-zero Horndeski potentials are now given by
\ba
\kappa_1&=&-\frac{3}{8}V_{paul}(\phi)\rho\\
\kappa_3&=&-\frac{1}{8}V'_{paul}(\phi)\rho
\ea
which give
\be
K_1=-\frac{3}{8}V_{paul}(\phi)\rho, \qquad K_3=-\frac{1}{8}V'_{paul}(\phi)
\ee
Again, using the expression (\ref{horndeskieom}), we find
\begin{multline} \label{eabpaul}
{\cal E}^{\eta \epsilon }_{paul}=\frac{3}{2} P^{\eta \mu \epsilon \nu} \rho \left( V_{paul}\nabla_{\mu} \nabla_\nu \phi +\frac{1}{3}V_{paul}' \nabla_\mu \phi \nabla_\nu \phi\right) \\ 
 +\half g^{\epsilon \theta} \delta^{\eta \alpha \beta \gamma}_{\theta \mu\nu\sigma} \left(V_{paul} \nabla^\mu\nabla_\alpha\phi\nabla^\nu\nabla_\beta\phi\nabla^\sigma\nabla_\gamma\phi +V_{paul}' \nabla^\mu \phi \nabla_\alpha\phi\nabla^\nu\nabla_\beta\phi\nabla^\sigma\nabla_\gamma\phi\right)
\end{multline}
One can check by direct, albeit non-trivial,  computation that these are the equations of motion obtained by variation of $\int d^4 x {\cal L}_{paul}$ where ${\cal L}_{paul}=\sqrt{-g} V_{paul}(\phi)P^{\mu\nu\alpha \beta} \nabla_\mu \phi \nabla_\alpha \phi \nabla_\nu \nabla_\beta \phi$. Note that equation \ref{eabpaul} may also be written more succintly,
\begin{multline}
{\cal E}^{\eta \epsilon }_{paul}=\frac{3}{2} P^{\eta \mu \epsilon \nu} \rho V_{paul}^{2/3}  \nabla_{\mu}  \left(V_{paul}^{1/3}  \nabla_\nu \phi \right)  
+\half g^{\epsilon \theta} \delta^{\eta \alpha \beta \gamma}_{\theta \mu\nu\sigma}  \nabla^\mu \left(V_{paul}^{1/3} \nabla_\alpha\phi\right) \nabla^\nu  \left(V_{paul}^{1/3} \nabla_\beta\phi\right) \nabla^\sigma \left(V_{paul}^{1/3} \nabla_\gamma\phi\right) 
\end{multline}
Moving on to George, we find that the non-vanishing Horndeski's potentials are
\beq
\kappa_9=-3V_{george}''\rho, \qquad F+2W=\frac{1}{2}V_{george}
\eeq
which gives
\be
K_9=V_{george}''\rho, \qquad {\cal F}+2{\cal W}=-\frac{1}{2}V_{george}
\ee
The resulting equation of motion is
\be
{\cal E}_{george}^{\eta \epsilon }= V_{george} G^{\eta \epsilon }+g^{\epsilon \theta} \delta^{\eta \alpha}_{\theta \mu}\left(V_{george}' \nabla_\alpha \nabla^\mu \phi+V_{george}'' \nabla_\alpha \phi \nabla^\mu \phi\right)
\ee
This is readily identified with the equations of motion obtained upon variation of $\int d^4 x {\cal L}_{george}$ where ${\cal L}_{george}=\sqrt{-g} V_{george}(\phi)R$. It may be written more succintly as
\be
{\cal E}_{george}^{\eta \epsilon }= V_{george} G^{\eta \epsilon }-(\nabla^\eta \nabla^\epsilon - g^{\eta \epsilon } \Box )V_{george}
\ee
Finally, we turn to Ringo. The non-zero potentials are given by
\ba
\kappa_1=2V_{ringo}'(\phi)\left(1+\frac{1}{2}\ln|\rho|\right),\qquad \kappa_3= V_{ringo}''(\phi)\ln|\rho|
\ea
At the level of the field equations (\ref{horndeskieom}) this means that 
\ba
\label{ringo1}
K_1=V_{ringo}',\qquad K_3=V_{ringo}''
\ea
The equations of motion now give 
\ba
{\cal E}^{\eta \epsilon }_{ringo}&=&-4 P^{\eta \mu \epsilon \nu} \left(V_{ringo}'\nabla_\mu \nabla_\nu \phi+V_{ringo}'' \nabla_\mu \phi \nabla_\nu \phi\right)
\ea
The equations of motion are recognised as those obtained in \cite{CDD} under metric variation of $\int d^4 x {\cal L}_{ringo}$ where ${\cal L}_{ringo}=\sqrt{-g}V_{ringo}(\phi) \GB$. Again, we may write it more succintly as
\ba
{\cal E}^{\eta \epsilon }_{ringo}&=&-4 P^{\eta \mu \epsilon \nu} \nabla_\mu \nabla_\nu V_{ringo}
\ea

\section{Radiative corrections about self-tuning vacua} \label{radiative}
To analyse the issue of radiative corrections to {\it the Fab Four}, we first need to choose a classical solution and identify the effective theory describing graviton and scalar fluctuations. Since we do not have a preferred cosmological solution at this stage, we shall restrict our attention to an heuristic analysis about a generic self-tuning vacuum, without specifying the form of the potentials. Our approach will be somewhat schematic  since the full system has a complicated tensor structure, and a more thorough analysis would represent an entire project of its own.  Nevertheless, we can still   obtain an order of magnitude estimate for the radiative corrections without paying too much attention to the particular tensor structure, signs, or the exact value of order one coefficients. To this end,  we write the {\it the Fab Four} Lagrangian schematically as follows:
\begin{multline}
L_{FabFour} \sim \sqrt{-g} \left[V_{john} (\phi) \nabla \phi \nabla \phi (\textrm{Einstein})+V_{paul}(\phi) \nabla \phi \nabla \phi  \nabla \nabla \phi (\textrm{P-tensor})\right. \\ \left.+V_{george}(\phi) R+V_{ringo}(\phi) \GB+\rho_\Lambda+\bar \psi(\dslash+m) \psi \right]
\end{multline}
where $\rho_\Lambda$ is the vacuum energy density. The matter coupling is represented by $\psi$, a fermion of mass, $m$. We neglect any subtleties involving the vierbein and coupling the spinor in curved space. Now let us expand the metric about the self-tuning Minkowski solution, $g_{\mu\nu}=\eta_{\mu\nu}+h_{\mu\nu}$. Schematically, we note that, 
\begin{equation}
\textrm{Einstein}, \textrm{P-tensor}, R \sim \sum_{n=1}^{\infty} (\partial)^2 h^n, \qquad \GB \sim \partial \cdot  \sum_{n=2}^{\infty} (\partial)^3 h^n, \qquad \sqrt{-g} \sim 1+\sum_{n=1}^{\infty} h^n
\end{equation}
Although we are obviously suppressing tensor structure, we are explicitly emphasizing the fact that in four dimensions, the Gauss-Bonnet combination is a total derivative, $\GB \sim \partial \cdot (\textrm{terms involving $h$})$. Thus our action can be written in the form,
\begin{multline}
L_{FabFour} \sim 
A(\phi, \partial \phi, \partial\partial \phi) \sum_{n=1}^{\infty} (\partial)^2 h^n +B(\phi, \partial \phi)  \sum_{n=2}^{\infty} (\partial)^3 h^n \\+\rho_\Lambda \left(1+\sum_{n=1}^{\infty} h^n\right)  +\bar \psi(\dslash+m) \psi  \left(1+\sum_{n=1}^{\infty} h^n\right)
\end{multline}
where $A \sim V_{john} (\phi) \partial \phi  \partial \phi +V_{paul}(\phi) \partial \phi \partial \phi \partial  \partial \phi+V_{george}(\phi) $, and $B \sim V_{ringo}'(\phi) \partial \phi $.
Now suppose that the background solution for the scalar is $\phi=\bar \phi(x)$. From the $h$ equation of motion we conclude that, $\partial \partial \bar A\sim \rho_\Lambda$, where ``bar"  denotes ``evaluated on the background"\footnote{For example,  $\bar A=A(\bar \phi, \partial \bar \phi, \partial\partial \bar \phi)$}. It follows that $\bar A \sim \rho_\Lambda x^2$.

We now consider fluctuations in $\phi$ of the form $\phi=\bar \phi+\xi$.  Working to lowest order in derivatives, we make the following low energy approximations
\begin{equation}
A \sim \bar A+\sum_{n=1}^\infty \frac{\partial^n A}{\partial \phi^n} \Big|_{\phi=\bar \phi}\xi^n, \qquad B \sim \bar B+\sum_{n=1}^\infty \frac{\partial^n B}{\partial \phi^n} \Big|_{\phi=\bar \phi} \xi^n
\end{equation}
This amounts to neglecting terms that go like, $p^{a_1+\cdots+a_N} \left[\left(\frac{\partial^N X}{\partial (\partial^{a_1} \phi \cdots \partial^{a_N} \phi)} \right)/ \left(\frac{\partial^N X}{\partial \phi^N}\right)\right]_{\phi=\bar \phi}$, where $X=A$ or $B$, and $p$ is momentum. Further assuming that $p \ll   \left[\frac{\partial^n A}{\partial \phi^n} /  \frac{\partial^n B}{\partial \phi^n} \right]_{\phi=\bar \phi}$, we find that up to cubic order in the fields, the effective Lagrangian has the following form in momentum space,
\be \label{effaction}
L_{eff}=K_{ij} q_i p^2 q_j +M_{ij} q_i q_j +\bar \psi (\pslash+m) \psi + \lambda_{ijk} q_i q_j q_k +n_i q_i \bar \psi (\pslash+m) \psi
\ee
where we define $q_1 \sim \sqrt{\bar A} h, ~q_2 \sim \frac{1}{ \sqrt{\bar A} }\frac{\partial A}{\partial \phi}\Big|_{\phi=\bar \phi} \xi$. The non-zero terms above are given by
\begin{multline}
 K_{11} \sim 1, \qquad K_{12} \sim 1,  \qquad M_{11} \sim \mu^2, \qquad n_1 \sim  \frac{1}{\sqrt{\bar A}}  \\ \lambda_{111} \sim \frac{1}{\sqrt{\bar A}} (p^2+\mu^2), \qquad 
  \lambda_{112} \sim \frac{1}{\sqrt{\bar A}} p^2, \qquad \lambda_{122} \sim \left[\frac{\frac{\partial^2 A}{\partial \phi^2} \sqrt{\bar A}}{\left(\frac{\partial A}{\partial \phi}\right)^2 }\right]_{\phi=\bar \phi}  p^2
 \end{multline}
 where we define $\mu^2 \sim \frac{\rho_{\Lambda}}{\bar A} \sim 1/x^2$, the latter relation following on from the fact that $\bar A \sim \rho_\Lambda x^2$.
 
 From now on, we will assume for simplicity that $\left[\frac{\frac{\partial^2 A}{\partial \phi^2} \sqrt{\bar A}}{\left(\frac{\partial A}{\partial \phi}\right)^2 }\right]_{\phi=\bar \phi}  \sim \frac{1}{\sqrt{\bar A}} $ in order that all the non-trivial three-point interactions involving $q_1$ and $q_2$ are of similar strength. Such behaviour is consistent with, say, exponential potentials. The theory defined by equation (\ref{effaction}) is only valid up to some momentum cut-off,  $\Lambda_{UV}$ (not to be confused with the cosmological constant!).   The form of (\ref{effaction}) suggests that  the classical interactions become strong at the scale $\sqrt{\bar A}$, and so we must at least have $\Lambda_{UV} \lesssim \sqrt{\bar A}$.   In any event,  we can only make sense of  the background on scales $x>\Lambda_{UV}^{-1}$. It follows that the mass scale $\mu<\Lambda_{UV}$, and if we further assume that $\Lambda_{UV} <\sqrt{\bar A}$ then we can ensure that the quantum interactions remain weakly coupled\footnote{Placing $\Lambda_{UV}$ {\it strictly} below $\sqrt{\bar A}$ amounts to saying that the UV completion of {\it the Fab Four} theory kicks in sooner than expected, and that these include irrelevant operators that already become important at energies of the order $\Lambda_{UV}$ when the classical interactions are still small. }.
 
 Let us now compute the one-loop correction to the bare Lagrangian (\ref{effaction}).  At tree level, the proper 2-vertices are given by $$\Gamma_{ij} \sim K_{ij}p^2 +M_{ij}, \qquad \Gamma_{\psi \bar\psi} \sim \pslash +m$$ and the proper 3-vertices by
 $$\Gamma_{ijk} \sim \lambda_{ijk}, \qquad \Gamma_{i \psi \bar \psi} \sim n_i (\pslash +m)$$
The tree-level propagators are just given by the inverse of the proper 2-vertices, 
$$G_{ij} = \Gamma^{ij}, \qquad G_{\psi \bar \psi} \sim \frac{1}{\pslash +m}  $$
where we denote the inverse with indices raised, $(\Gamma^{-1})_{ij}=\Gamma^{ij}$. We immediately note that $G_{11}=0$, while $G_{12} \sim G_{22} \sim 1$. This means that we have no $h-h$ propagator at tree level.

To compute the one loop correction to the propagator, $G_{ij}$, we will need knowledge of the self energy, $\Sigma_{ij}$ at one loop. Let us postpone this until later. For the moment, let us concentrate on summing up the relevant 1PI graphs. The renormalised propagator is given by
\begin{eqnarray}
&& G^{ren}_{ij} = G_{ij}+G_{ik} \Sigma_{kl} G_{lj} +G_{ik} \Sigma_{kl} G_{lm} \Sigma_{mn} G_{nj} +\ldots \\
&\implies& G^{ren}=G(1-\Sigma G)^{-1} 
\end{eqnarray}
It follows that the renormalised proper 2-vertex is given by $\Gamma_{ij}^{ren}=(G^{ren})_{ij}^{-1}=\Gamma_{ij}-\Sigma_{ij}$.

\begin{figure}[t] 
  \centering
  \includegraphics[width=4in]{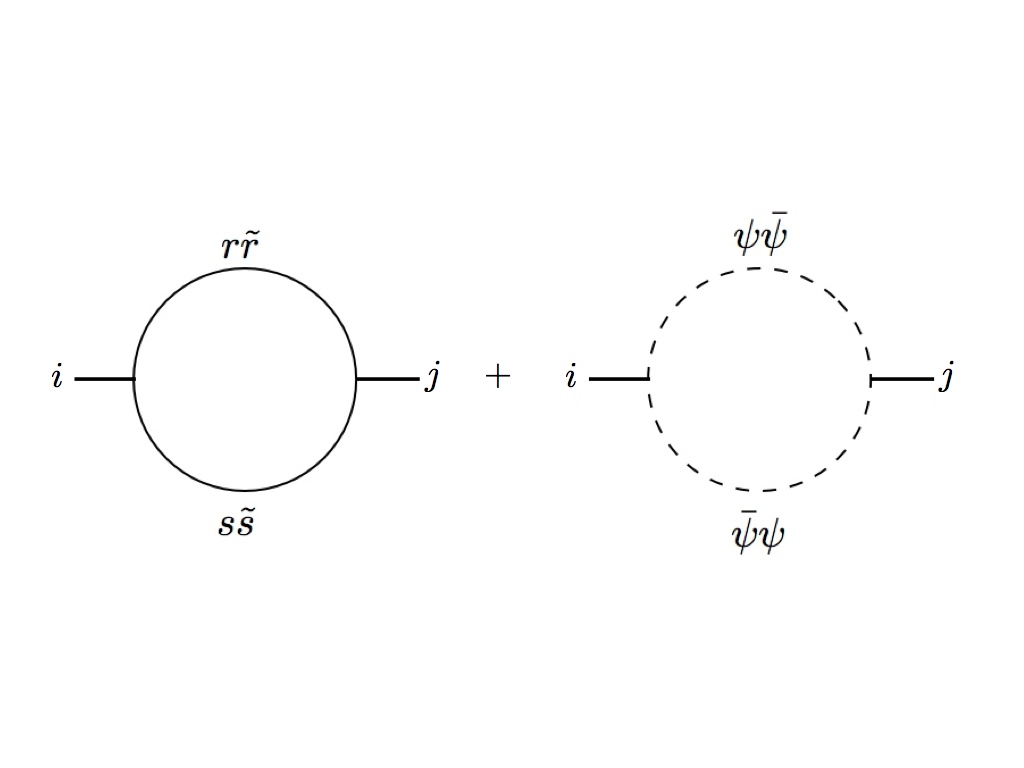} 
  \caption{Feynman diagrams for $\Sigma_{ij}$.}
    \label{graphs1}
\end{figure}

We now compute $\Sigma_{ij}$. The relevant graphs are shown in figure \ref{graphs1}.
We find that
\begin{equation}
\Sigma_{ij} \sim \lambda_{irs}\lambda_{j \tilde r\tilde s} \int d^4 k G_{r\tilde r}(k) G_{s\tilde s}(p-k)+n_i n_j (\pslash +m)^2 \int d^4 k G_{\psi \bar \psi} (k) G_{\bar \psi \psi}(p-k)
\end{equation}
Now, since $\mu< \Lambda_{UV}$, we find  that $\int d^4 k G_{r\tilde r}(k) G_{s\tilde s}(p-k) \sim K^{r \tilde r} K^{s\tilde s} \log (\Lambda_{UV}/\mu)$, while 
\begin{equation}
\int d^4 k G_{\psi \bar \psi} (k) G_{\bar \psi \psi}(p-k)=I(\Lambda_{UV}) =\begin{cases} \Lambda_{UV}^2 & m< \Lambda_{UV} \\ \frac{\Lambda_{UV}^4}{m^2} & m>\Lambda_{UV}\end{cases}\end{equation}
Note that $I(\Lambda_{UV})  \lesssim \Lambda_{UV}^2$ and $I(\Lambda_{UV})  \lesssim \Lambda_{UV}^4/m^2$. After some calculation, we can further show that 
\begin{equation}
\Sigma_{ij} \sim \frac{p^4}{\bar A} \log (\Lambda_{UV}/\mu)+ \delta_{1i} \delta_{1j} \left[ \frac{p^2\mu^2}{\bar A}  \log (\Lambda_{UV}/\mu)+\frac{(\pslash +m)^2}{\bar A} I(\Lambda_{UV}) \right]
\end{equation}
Let us use this to compute the one loop corrections to $K_{ij}$ and $M_{ij}$.  For $p>m$, we have $ \Delta M_{ij} \approx 0$ and 
\begin{equation}
\Delta K_{ij} \sim p^2 \left[\frac{p^2}{\bar A} \log (\Lambda_{UV}/\mu)+ \delta_{1i} \delta_{1j} \left( \frac{\mu^2}{\bar A}  \log (\Lambda_{UV}/\mu)+\frac{\Lambda_{UV}^2}{\bar A} \frac{I(\Lambda_{UV})}{\Lambda_{UV}^2}\right)  \right].
\end{equation}
Since $p^2, \mu^2,  I(\Lambda_{UV})  \lesssim \Lambda_{UV}^2$, it is clear that $\Delta K_{ij} < K_{ij}$ whenever $\bar A> \Lambda_{UV}^2$.

For $p<m$ the situation is slightly different. Then we find that 
\begin{equation}
\Delta M_{ij} \sim  \delta_{1i} \delta_{1j} \frac{m^2 I( \Lambda_{UV})}{\bar A} , \qquad 
\Delta K_{ij} \sim p^2 \left[\frac{p^2}{\bar A} \log (\Lambda_{UV}/\mu)+ \delta_{1i} \delta_{1j} \left( \frac{\mu^2}{\bar A}  \log (\Lambda_{UV}/\mu) \right)\right]
\end{equation}
As before, it is sufficient to take $\bar A> \Lambda_{UV}^2$ to ensure that $\Delta K_{ij} < K_{ij}$.  We now compare $\Delta M_{ij}$  with $M_{ij}$, noting that
\begin{equation}
\frac{m^2 I(\Lambda_{UV})}{\mu^2 \bar A}   \sim \frac{m^2 I(\Lambda_{UV})}{\rho_\Lambda}  \lesssim 
\frac{\Lambda_{UV}^4}{\rho_\Lambda} 
\end{equation}
where we have used the fact that $I(\Lambda_{UV})  \lesssim \Lambda_{UV}^4/m^2$ and $\mu^2 \sim \rho_\Lambda/\bar A$. It now follows that $\Delta M_{ij} <M_{ij}$ if we take $\Lambda_{UV} < \rho_\Lambda^{1/4}$.

We therefore conclude that one-loop corrections to $K_{ij}$ and $M_{ij}$ are suppressed provided we take $\Lambda_{UV} < \sqrt{\bar A}, \rho_\Lambda^{1/4}$. Indeed, we have also checked that these conditions also ensure that one-loop corrections to the 3 vertices, $\lambda_{ijk}$ are also suppressed. We are almost done. However, it is important to realise that our analysis also implies a {\it lower} bound on $\Lambda_{UV}$. This is because $\sqrt{\bar A} \sim x \sqrt{\rho_\Lambda} >\sqrt{\rho_\Lambda}/\Lambda_{UV}$, and so we have $\Lambda_{UV}>\sqrt{\rho_\Lambda/\bar A}$. All necessary conditions may be encapsulated in the following statement,
\begin{equation}
\sqrt{{G}_\textrm{eff}\rho_\Lambda}<\Lambda_{UV} < \rho_\Lambda^{1/4}
\end{equation}
Here we have identified ${G}_\textrm{eff} \sim 1/\bar A$, as the (time dependent) strength of the gravitational coupling to matter, in the linearised regime (that is, neglecting any possible Vainshtein effects etc etc).


\begin{thebibliography}{99}
\bibitem{Charmousis:2011bf}
  C.~Charmousis, E.~J.~Copeland, A.~Padilla, P.~M.~Saffin,
   [arXiv:1106.2000 [hep-th]].

\bibitem{horndeski:1974}
  G.~W.~Horndeski,
  Int.\ J.\ Theor.\ Phys.\  {\bf 10}, 363 (1974).

\bibitem{review}
  T.~Clifton, P.~G.~Ferreira, A.~Padilla, C.~Skordis,
  [arXiv:1106.2476 [astro-ph.CO]].
  
\bibitem{general}
  C.~Deffayet, X.~Gao, D.~A.~Steer, G.~Zahariade,
    [arXiv:1103.3260 [hep-th]].


\bibitem{bdgravity}
  C.~Brans, R.~H.~Dicke,
  Phys.\ Rev.\  {\bf 124 } (1961)  925-935.

\bibitem{kkl}

  F.~Mueller-Hoissen,
  Class.\ Quant.\ Grav.\  {\bf 3}, 665 (1986).
  F.~Mueller-Hoissen,
  Phys.\ Lett.\ B {\bf 201} (1988) 325.
  F.~Mueller-Hoissen,
  Nucl.\ Phys.\ B {\bf 337}, 709 (1990).
  C.~Cartier, J.~-c.~Hwang and E.~J.~Copeland,
  Phys.\ Rev.\ D {\bf 64}, 103504 (2001)
  [astro-ph/0106197].
  L.~Amendola, C.~Charmousis and S.~C.~Davis,
  JCAP {\bf 0612}, 020 (2006)
  [hep-th/0506137].


\bibitem{covgal}
  C.~Deffayet, G.~Esposito-Farese, A.~Vikman,
  Phys.\ Rev.\  {\bf D79 } (2009)  084003.
  [arXiv:0901.1314 [hep-th]].
  
\bibitem{galmodels}
  C.~Deffayet, O.~Pujolas, I.~Sawicki, A.~Vikman,
  JCAP {\bf 1010 } (2010)  026.
  [arXiv:1008.0048 [hep-th]].
  F.~P.~Silva, K.~Koyama,
  Phys.\ Rev.\  {\bf D80 } (2009)  121301.
  [arXiv:0909.4538 [astro-ph.CO]].  
  T.~Kobayashi, M.~Yamaguchi, J.~'i.~Yokoyama,
  Phys.\ Rev.\ Lett.\  {\bf 105 } (2010)  231302.
  [arXiv:1008.0603 [hep-th]].  
  A.~De Felice, S.~Tsujikawa,
   [arXiv:1008.4236 [hep-th]].   
  C.~Deffayet, S.~Deser and G.~Esposito-Farese,
  Phys.\ Rev.\  D {\bf 80}, 064015 (2009)
  [arXiv:0906.1967].
  

\bibitem{galileon}
  A.~Nicolis, R.~Rattazzi, E.~Trincherini,
  Phys.\ Rev.\  {\bf D79 } (2009)  064036.
  [arXiv:0811.2197 [hep-th]].

\bibitem{VanAcoleyen:2011mj} 
  K.~Van Acoleyen and J.~Van Doorsselaere,
  Phys.\ Rev.\ D {\bf 83}, 084025 (2011)
  [arXiv:1102.0487 [gr-qc]].



\bibitem{Kobayashi:2011nu}
  T.~Kobayashi, M.~Yamaguchi, J.~'i.~Yokoyama,
  [arXiv:1105.5723 [hep-th]].
  


  
\bibitem{DeFelice:2011hq}
  A.~De Felice, T.~Kobayashi, S.~Tsujikawa,
  [arXiv:1108.4242 [gr-qc]].
  
 \bibitem{ostro} 
  M. ~Ostrogradsky, 
  Memoires de l�Academie Imperiale des Science de Saint-Petersbourg, 4:385, 
1850. 
  
    
\bibitem{nogo}
  S.~Weinberg,
  Rev.\ Mod.\ Phys.\  {\bf 61 } (1989)  1-23.
  
\bibitem{bigalileon}
  A.~Padilla, P.~M.~Saffin, S.~-Y.~Zhou,
  JHEP {\bf 1012 } (2010)  031.
  [arXiv:1007.5424 [hep-th]].
  A.~Padilla, P.~M.~Saffin, S.~-Y.~Zhou,
  JHEP {\bf 1101 } (2011)  099.
  [arXiv:1008.3312 [hep-th]].

\bibitem{mtw}
  C.~W.~Misner, K.~S.~Thorne, J.~A.~Wheeler,
  San Francisco 1973, 1279p.
  
  
   \bibitem{fab4cos}
  C.~Charmousis, E.~J.~Copeland, A.~Padilla, P.~M.~Saffin,
  ``The cosmology of the Fab Four,''

  
  
\bibitem{Carroll}
  S.~M.~Carroll, M.~M.~Guica,
  [hep-th/0302067].
  
\bibitem{cline}
  J.~Vinet, J.~M.~Cline,
  Phys.\ Rev.\  {\bf D70 } (2004)  083514.
  [hep-th/0406141].
  H.~-P.~Nilles, A.~Papazoglou, G.~Tasinato,
  Nucl.\ Phys.\  {\bf B677 } (2004)  405-429.
  [arXiv:hep-th/0309042 [hep-th]].
  
\bibitem{Amendola:2008vd}
  L.~Amendola, C.~Charmousis and S.~C.~Davis,
  Phys.\ Rev.\  D {\bf 78}, 084009 (2008)
  [arXiv:0801.4339 [gr-qc]].

\bibitem{vainsh}
  A.~I.~Vainshtein,
  Phys.\ Lett.\  {\bf B39 } (1972)  393-394.
   
\bibitem{cham}
  J.~Khoury, A.~Weltman,
  Phys.\ Rev.\ Lett.\  {\bf 93 } (2004)  171104.
  [astro-ph/0309300].


\bibitem{lanczos}
  C.~Lanczos,
  Rev.\ Mod.\ Phys.\  {\bf 34}, 379-389 (1962).

\bibitem{gravitation}
Misner, Charles W.; Thorne, Kip S.; Wheeler, John Archibald (1973), Gravitation, San Francisco


\bibitem{CDD}
  C.~Charmousis, S.~C.~Davis and J.~F.~Dufaux,
  JHEP {\bf 0312}, 029 (2003)
  [arXiv:hep-th/0309083].

\bibitem{stephen}
  S.~C.~Davis,
  AIP Conf.\ Proc.\  {\bf 736}, 147-152 (2005).
  [hep-th/0410075].


\bibitem{lovelock}
  D.~Lovelock,
  J.\ Math.\ Phys.\  {\bf 12}, 498-501 (1971)


\bibitem{cc2}
  C.~Charmousis,
  Lect.\ Notes Phys.\  {\bf 769}, 299-346 (2009).
  [arXiv:0805.0568 [gr-qc]].
  
\bibitem{cosmogal}
  C.~de Rham and L.~Heisenberg,
  Phys.\ Rev.\ D\ {\bf 84} (2011) 043503
  [arXiv:1106.3312 [hep-th]].

\bibitem{dgpvainsh}
  C.~Deffayet, G.~R.~Dvali, G.~Gabadadze and A.~I.~Vainshtein,
  Phys.\ Rev.\ D\ {\bf 65} (2002) 044026
  [hep-th/0106001].
  
\bibitem{hirsute}
  N.~Kaloper, A.~Padilla and N.~Tanahashi,
  JHEP\ {\bf 1110} (2011) 148
  [arXiv:1106.4827 [hep-th]].
  
\bibitem{vainshgal}
  A.~De Felice, R.~Kase and S.~Tsujikawa,
  arXiv:1111.5090 [gr-qc].
  
\bibitem{higgs}
  C.~Germani and A.~Kehagias,
  Phys.\ Rev.\ Lett.\ \ {\bf 105} (2010) 011302
  [arXiv:1003.2635 [hep-ph]].


\end{thebibliography}
\end{document}